\documentclass[reprint,nofootinbib,aps,prd,superscriptaddress,floatfix]{revtex4-2}
\usepackage{amsmath,amssymb,mathrsfs,bm,graphicx,booktabs,xcolor,hyperref}
\usepackage{orcidlink}
\hypersetup{colorlinks=true,linkcolor=blue,citecolor=blue,urlcolor=blue}
\newcommand{\dd}{\mathrm d}
\newcommand{\Mpl}{M_{\rm Pl}}
\newcommand{\HI}{H_I}
\newcommand{\PS}{\mathcal P_S}
\newcommand{\cRD}{c_{\rm RD}}
\newcommand{\Iprod}{\mathcal I}
\newcommand{\Ftr}{\mathscr F}
\newcommand{\GeV}{\,\mathrm{GeV}}
\newcommand{\Mpc}{\,\mathrm{Mpc}^{-1}}
\graphicspath{{figures/}}

\begin{document}
\raggedbottom

\title{Can inflationary axion fluctuations dominate over misalignment?}

\author{Seung J. Lee\,\orcidlink{0000-0002-7756-0407}}
\email{sjjlee@kias.re.kr}
\affiliation{School of Physics and Quantum Universe Center, Korea Institute for Advanced Study, Seoul 02455, Korea}
\author{Sarunas Verner\,\orcidlink{0000-0003-4870-0826}}
\email{verner@uchicago.edu}
\affiliation{Kavli Institute for Cosmological Physics, University of Chicago, Chicago, Illinois 60637, USA}
\author{Bingrong Yu\,\orcidlink{0000-0001-5842-6563}}
\email{bingrong.yu@cornell.edu}
\affiliation{Department of Physics, LEPP, Cornell University, Ithaca, New York 14853, USA}

\date{\today}

\begin{abstract}
The relic abundance of light stable axion dark matter is usually set by misalignment. We study whether the inflationary fluctuations of the axion can dominate instead, and we follow the modes analytically through an instantaneous transition from de~Sitter expansion to radiation domination, neglecting the axion mass during inflation and keeping it afterward.
Since the energy density is quadratic in the field, a fluctuation-dominated abundance carries an isocurvature perturbation even when the mean displacement vanishes. For a canonical minimally coupled axion the spectrum is scale invariant, and fluctuation-dominated dark matter exceeds the Planck isocurvature limit by about eight orders of magnitude. An inflaton-dependent kinetic function or a nonminimal coupling to the Ricci scalar tilts the spectrum blue, moving the fluctuation power from CMB scales to small scales. Imposing the Planck limit at the pivot scale, we find that fluctuation-dominated dark matter with a constant kinetic slope requires the spectral tilt $n_\eta-1\gtrsim0.27$ and the axion mass $m_\eta\gtrsim200\GeV$ at $\HI=10^{10}\GeV$, and $n_\eta-1\gtrsim0.34$ and $m_\eta\gtrsim4.5\,$keV at $\HI=10^{13}\GeV$. A kinetic function with a running slope lowers these masses to $1.4\GeV$ and $0.7\,$eV, respectively. For the QCD axion, whose oscillations start late, we estimate that any single power-law spectrum blue enough to satisfy the pivot limit gives too little fluctuation abundance, and misalignment dominates.

\end{abstract}

\maketitle

\section{Introduction}
\label{sec:intro}
The axion is a well-motivated dark matter candidate, and its relic abundance is usually set by the misalignment mechanism~\cite{Preskill:1982cy,Abbott:1982af,Dine:1982ah}, in which Hubble friction holds a homogeneous displacement of the field fixed until the Hubble rate drops below the axion mass, after which the field oscillates and redshifts like matter. If the Peccei--Quinn (PQ) symmetry~\cite{Peccei:1977hh, Peccei:1977ur, Weinberg:1977ma,Wilczek:1977pj} is broken before the observable scales leave the horizon, the axion is also a light spectator during inflation and acquires fluctuations of order $\HI/(2\pi)$, where $\HI$ is the inflationary Hubble rate. These fluctuations were first studied as a source of axion density perturbations in Ref.~\cite{Axenides:1983hj} and soon identified as isocurvature perturbations~\cite{Seckel:1985tj,Linde:1985yf}. The absence of isocurvature in the cosmic microwave background (CMB) bounds the inflationary Hubble scale~\cite{Lyth:1989pb,Turner:1990uz,Linde:1991km,Lyth:1992tx,Fox:2004kb, Hertzberg:2008wr, Planck:2018jri}, and the bound depends on the axion decay constant during inflation, on the later evolution of the perturbation, and on the ultraviolet (UV) completion~\cite{Linde:1991km,Kearney:2016vqw,Graham:2025iwx,Berbig:2024ufe,Verner:2026cbn}.

The same fluctuations carry energy and contribute to the abundance. Their contribution can be computed either as gravitational particle production~\cite{Parker:1969au,Ford:1986sy,Birrell:1982ix,Kaneta:2022gug} (see Ref.~\cite{Kolb:2023ydq} for a review) or as the classical evolution of a stochastic field. Fields with masses near the inflationary Hubble rate are produced mainly around the end of inflation~\cite{Chung:1998zb,Ema:2018ucl,Ling:2021zlj,Racco:2024aac,Verner:2024agh}. For light scalars, on the other hand, the infrared modes generate large isocurvature, which a nonminimal coupling to gravity can suppress~\cite{Garcia:2023awt,Garcia:2023qab}. Ref.~\cite{Herring:2019hbe} computed the transition from de~Sitter expansion to radiation domination analytically for a light minimally coupled spectator and found a late-time distribution strongly peaked at low momenta. If inflation lasts long enough, a light field approaches the stochastic equilibrium distribution~\cite{Starobinsky:1994bd}, whose variance can replace the initial displacement~\cite{Graham:2018jyp,Takahashi:2018tdu}. Related ideas have been applied to spectator scalar dark matter~\cite{Markkanen:2018gcw,Tenkanen:2019aij, Nurmi:2015ema, Garcia:2025rut} and to vector dark matter~\cite{Graham:2015rva,Ema:2019yrd,Ahmed:2020fhc,Kolb:2020fwh}. 

A time-dependent kinetic function also tilts the axion spectrum. Such kinetic functions were introduced for inflationary magnetogenesis~\cite{Ratra:1991bn,Martin:2007ue}, dark photon production~\cite{Nakai:2020cfw,Nakayama:2020rka}, and axion dark matter production~\cite{Ismail:2024zbq}. For an axion, the kinetic function can come from an evolving radial expectation value of the PQ field~\cite{Linde:1991km,Kearney:2016vqw,Graham:2025iwx} or from a curved field-space metric (e.g., in supergravity and string compactifications~\cite{Cremmer:1982en,Svrcek:2006yi,Tadepalli:2026mdc}).

The CMB bound can be avoided either by suppressing the isocurvature or by tilting it to become more blue. A larger effective decay constant during inflation suppresses the axion isocurvature, either through a large radial field~\cite{Linde:1991km,Kearney:2016vqw,Graham:2025iwx,Kobayashi:2016qld} or through the field-space metric~\cite{Tadepalli:2026mdc}, and a nonminimal kinetic coupling to gravity has a similar effect~\cite{Folkerts:2013tua}. A large axion mass during inflation also removes the CMB scale isocurvature~\cite{Dvali:1995ce,Jeong:2013xta,Higaki:2014ooa}, and if the axion becomes light only after the CMB scales exit, the later fluctuations can set the QCD axion abundance~\cite{Kitano:2023mra,Freese:2026xax}. An extremely blue spectrum arises instead if the radial field of the axion evolves during inflation~\cite{Kasuya:2009up,Chung:2015pga,Chung:2016wvv}. For a self-interacting scalar, whose condensate spends most of inflation near the boundary of the slow-roll region of its potential, the effective mass stays close to $\HI$ and the tilt is generated dynamically~\cite{Kalia:2025uxg}. A positive coupling to the Ricci scalar gives a blue tilt and erases the initial displacement, which allows scalar dark matter to come entirely from inflationary fluctuations~\cite{Alonso-Alvarez:2018tus,Alonso-Alvarez:2019ixv}. The two couplings that we study here, a kinetic function and a curvature coupling, act in both ways.

In this work, we ask when inflationary fluctuations dominate the abundance of an axion-like particle (ALP) compared to the conventional misalignment production and what this implies for isocurvature. We extend the analytic matching of Ref.~\cite{Herring:2019hbe} to scenarios with either a nonminimal coupling or a noncanonical kinetic term, both of which have well-motivated UV realizations. 
Our matching keeps the axion mass after inflation and the jump in the rate of change of the kinetic function at the end of inflation, and gives the late-time spectrum fully analytically for both couplings. The same calculation fixes the misalignment normalization without an onset prescription and shows that the fluctuations carry the misalignment abundance of an effective homogeneous displacement, so that the two contributions can be compared directly. 

The isocurvature perturbation contains a term linear in the fluctuations, proportional to the mean displacement, which reproduces the familiar misalignment result, and a term quadratic in the fluctuations that remains when the mean vanishes. 
We then confront the fluctuation-dominated abundance with the current CMB isocurvature limit, including the quadratic term that survives when the mean vanishes. 
We identify the conditions under which axions produced from inflationary fluctuations can dominate over misalignment while satisfying the stringent isocurvature constraint. The early estimate of Ref.~\cite{Turner:1987vd} for the QCD axion predates this constraint, and the red spectra that give a large abundance there exceed it at the pivot by many orders of magnitude (see Sec.~\ref{sec:iso}). The more comprehensive treatment here also revises the conclusion of Ref.~\cite{Ismail:2024zbq} that red spectra can supply all of the dark matter. That analysis imposed the isocurvature bound in its misalignment-dominated form (the linear term), which misses the quadratic term that dominates once the fluctuations set the abundance. 


For the noncanonical kinetic coupling, we consider two scenarios. A linear exponential kinetic function realizes the constant-slope history that we solve exactly, whereas a quadratic exponential produces a running slope, with a spectrum that is strongly blue on CMB scales and less blue on the smaller scales that dominate the abundance. Since these scales leave the horizon tens of $e$-folds apart, running can suppress the CMB-scale isocurvature relative to the fluctuation abundance, substantially relaxing the mass thresholds obtained with a constant slope. Inflationary fluctuations can then supply all of the dark matter for axion masses as low as $0.7\,\mathrm{eV}$ at $\HI=10^{13}\GeV$ and approximately $0.01\,\mathrm{eV}$ at the tensor bound, $\HI=4.7\times10^{13}\GeV$, while satisfying the pivot-scale isocurvature constraint.

The remainder of the paper is organized as follows. In Sec.~\ref{sec:setup}, we define the model, and in Sec.~\ref{sec:production} we solve the production problem fully analytically. We compute the abundance in Sec.~\ref{sec:abundance} and the isocurvature in Sec.~\ref{sec:iso}, which also contains an estimate for the QCD axion (Sec.~\ref{sec:QCD}). In Sec.~\ref{sec:slopes}, we relate the analytic slope to the linear and quadratic exponentials, including the running of the slope, and in Sec.~\ref{sec:conclusions} 
we summarize our results. Apps.~\ref{app:IR}--\ref{app:histories} derive the infrared spectrum and the density correlator and treat slowly varying kinetic histories, and App.~\ref{app:UV} gives UV realizations of the kinetic functions.

\section{Setup}
\label{sec:setup}
We consider a canonical inflaton $\phi$ and a spectator axion $\eta$ with action
\begin{equation}
\begin{aligned}
S=\int\dd^4x\sqrt{-g}\,\bigg[&\frac{\Mpl^2}{2}R-\frac12(\partial\phi)^2-V(\phi)\\
&-\frac12K^2(\phi)(\partial\eta)^2-\frac12\xi R\eta^2-U(\eta)\bigg],
\end{aligned}
\label{eq:action}
\end{equation}
where $\Mpl=2.435\times10^{18}\GeV$ is the reduced Planck mass, $R$ is the Ricci scalar, and the signature is $(-,+,+,+)$. The functions $V$ and $U$ are the inflaton and axion potentials, $K^2(\phi)>0$ is the axion kinetic metric, and $\xi$ is dimensionless. We study the two couplings separately, either a varying $K(\phi)$ with $\xi=0$~\cite{Ismail:2024zbq} or a constant $\xi$ with $K=1$~\cite{Alonso-Alvarez:2018tus,Alonso-Alvarez:2019ixv}, and we take the canonical minimally coupled theory ($K=1$ and $\xi=0$) as our reference case. We work in a spatially flat FLRW background, $\dd s^2=-\dd t^2+a^2\dd\bm x^2=a^2(-\dd\tau^2+\dd\bm x^2)$, with Hubble rate $H=\dot a/a$, where dots and primes denote derivatives with respect to cosmic time $t$ and conformal time $\tau$, respectively. With these conventions, the Ricci scalar is given by
\begin{equation}
R=6(2H^2+\dot H)=\frac{6a''}{a^3}.
\label{eq:Ricci}
\end{equation}

We use the term ``axion'' for any light pseudoscalar with an approximate shift symmetry. One realization is the angular mode of a complex PQ field,
\begin{equation}
\chi=\frac{\sigma}{\sqrt2}\,e^{i\eta/f_\eta},
\qquad
\eta\sim\eta+2\pi f_\eta,
\label{eq:PQfield}
\end{equation}
where $\sigma$ is the radial field and $f_\eta$ is the late-time axion decay constant. We assume that PQ symmetry is broken before observable modes leave the horizon and stays broken, and that the radial field tracks a stable minimum $\sigma_0(\phi)$ with mass $m_\sigma$ well above the inflationary Hubble rate $\HI$, the relevant physical momenta, and the rates of background evolution.\footnote{A PQ-invariant coupling $R|\chi|^2$ can shift the radial expectation value and hence the axion kinetic normalization~\cite{Graham:2025iwx}. The angular interaction $\xi R\eta^2$ instead requires explicit PQ breaking~\cite{Berbig:2024ufe}.} Integrating out $\sigma$ of a canonical complex field gives the axion kinetic function $K=\sigma_0/f_\eta$~\cite{Linde:1991km,Kearney:2016vqw,Graham:2025iwx}, while a curved field-space metric provides another realization (App.~\ref{app:sugra}).

The symmetry must also stay broken after inflation, which for instantaneous reheating, with a radiation temperature $T_{\rm RH} \simeq8\times10^{13}\GeV\,(\HI/10^{10}\GeV)^{1/2}$, requires $f_{\eta} \gtrsim T_{\rm RH}$ if the PQ sector is in equilibrium with the bath. However, a PQ sector that is coupled weakly enough never to thermalize is not restored even for $f_\eta<T_{\rm RH}$.\footnote{Nonthermal restoration through parametric resonance of $\sigma$ must be checked separately. Our results depend on $f_\eta$ only through the initial misalignment angle $\theta_c=\eta_c/f_\eta$, the harmonic conditions, and the photon coupling used for illustration.}

We take $K(\phi_e)=1$ at the end of inflation and assume that the exit keeps $K=1$ thereafter, i.e., $\eta$ is canonically normalized after reheating. At fixed $\phi$ the canonical angular field has period $2\pi K(\phi)f_\eta$, and its decay constant at horizon exit of the CMB pivot scale, $k_*=a_*H_*$, is given by
\begin{equation}
f_I\equiv K_*f_\eta,\qquad K_*\equiv K(\phi_*).
\label{eq:fI}
\end{equation}

The kinetic coupling preserves the shift symmetry, while $U$ and the curvature interaction break it. The latter can arise, for a compact axion, from the periodic operator $-\xi Rf_\eta^2[1-\cos(\eta/f_\eta)]$. We expand both interactions about $\eta=0$ and work in the harmonic regime, $U(\eta)\simeq\frac12m_\eta^2\eta^2$, with constant $m_\eta^2\equiv U_{,\eta\eta}(0)$. During inflation we retain the curvature mass but neglect the potential mass, which requires $m_\eta\ll K\HI$ along the relevant part of the trajectory.\footnote{The axion is a harmonic spectator if $|\rho_\eta|\ll3\Mpl^2H^2$, $|\xi|\langle\eta^2\rangle\ll\Mpl^2$, and $\langle\eta^2\rangle\ll f_\eta^2$, where $\rho_\eta$ includes the curvature term and the vacuum-subtracted $\langle\eta^2\rangle$ includes the mean displacement. For varying $K$, the term $KK_{,\phi}(\partial\eta)^2$ in the inflaton equation must also be negligible. For a dynamical decay constant the last condition also excludes small-scale fluctuations that exceed the period of the potential, which would form domain walls~\cite{Kobayashi:2016qld}. We treat the temperature-dependent mass of the QCD axion separately in Sec.~\ref{sec:QCD}.}

We consider kinetic functions whose logarithm is linear or quadratic in $\phi$,
\begin{equation}
K_\lambda^2(\phi)=e^{\lambda\phi/\Mpl},
\qquad
K_\gamma^2(\phi)=e^{\gamma\phi^2/\Mpl^2},
\label{eq:Kfunctions}
\end{equation}
with dimensionless $\lambda$ and $\gamma$, which we call the linear and quadratic exponentials. The quadratic exponential starts with the dimension-six interaction $-(\gamma/2\Mpl^2)\phi^2(\partial\eta)^2$, and its higher terms allow a larger slope $|p|$ than the truncated form (App.~\ref{app:radial}). Both functions equal unity at $\phi=0$, and if inflation ends at $\phi_e\neq0$, the conditions $K(\phi_e)=1$ and $K=1$ thereafter require modifying them near the exit.

We characterize the kinetic evolution by
\begin{equation}
p\equiv\frac{\dd\ln K}{\dd\ln a}
=\frac{K_{,\phi}}{K}\frac{\dot\phi}{H},
\label{eq:pdef}
\end{equation}
where $K_{,\phi}\equiv\dd K/\dd\phi$, and a positive $p$ corresponds to a growing $K$. For a canonical inflaton rolling toward smaller $\phi$ we have $\dot\phi/H\simeq-\sqrt{2\epsilon}\,\Mpl$, with $\epsilon\equiv-\dot H/H^2$ being the slow-roll parameter, which gives
\begin{equation}
p_\lambda=-\lambda\sqrt{\frac{\epsilon}{2}},
\qquad
p_\gamma=-\sqrt{2\epsilon}\,\gamma\,\frac{\phi}{\Mpl}.
\label{eq:plambdagamma}
\end{equation}
The blue spectra considered below correspond to $p<0$, and hence to $\lambda>0$ or, on the $\phi>0$ branch, to $\gamma>0$. Note that the linear exponential gives a constant $p$ when $\epsilon$ is constant, while $|p_\gamma|$ decreases as $\phi$ approaches zero.

Our calculations in Secs.~\ref{sec:production}-\ref{sec:iso} only assume a constant $p$ and do not rely on the specific form of $K(\phi)$. We map our results onto the specific forms in Eq.~(\ref{eq:Kfunctions}) in Sec.~\ref{sec:slopes}.

\section{Gravitational axion production}
\label{sec:production}
With the canonical variable $v\equiv z\eta$, where $z\equiv aK$, and with the potential mass neglected, the quadratic action during inflation can be written, after an integration by parts, as
\begin{equation}
\mathcal{S}_{\eta}=\frac12\int\dd\tau\,\dd^3x
\left[v'^2-(\bm\nabla v)^2-\mathcal M^2v^2\right],
\label{eq:canonicalaction}
\end{equation}
where
\begin{equation}
\mathcal M^2=
\begin{cases}
-\dfrac{(aK)''}{aK}, & \xi=0,\\[8pt]
(6\xi-1)\dfrac{a''}{a}, & K=1.
\end{cases}
\label{eq:canonicalmass}
\end{equation}
The modes of comoving wavenumber $k$ obey
\begin{equation}
v_k''+\left(k^2+\mathcal M^2\right)v_k=0,
\qquad
v_kv_k^{\prime*}-v_k^*v_k'=i.
\label{eq:mode}
\end{equation}

We approximate reheating as an instantaneous transition from de~Sitter expansion to radiation domination~\cite{Herring:2019hbe}. Setting $a_e=1$ and $\tau_e=-\HI^{-1}$, the scale factor is given by
\begin{equation}
a(\tau)=
\begin{cases}
-(\HI\tau)^{-1}, & \tau\leq\tau_e,\\
2+\HI\tau, & \tau\geq\tau_e.
\end{cases}
\label{eq:background}
\end{equation}
Both $a$ and $a'$ are continuous at $\tau_e$, and $R$ has no delta function there. Instead, it drops from $12\HI^2$ to zero, and the curvature mass switches off. In the kinetic coupling case, we take   
\begin{equation}
K(\tau)=
\begin{cases}
a^p(\tau), & \tau<\tau_e,\\
1, & \tau\geq\tau_e,
\end{cases}
\label{eq:powerK}
\end{equation}
with a constant $p$. This treats the geometry at leading order in slow roll while retaining the evolution of $K$ (we relate it to Eq.~\eqref{eq:Kfunctions} in Sec.~\ref{sec:slopes}). The kinetic function is continuous at the transition, where $K'$ drops from $p\HI$ to zero, and for the curvature coupling $K=1$ throughout.

In de~Sitter space, $(aK)''/(aK)=(2+\kappa)/\tau^2$, where
\begin{equation}
\kappa\equiv\tau^2\left[
\frac{K''}{K}+2\frac{a'}{a}\frac{K'}{K}
\right]
=p(p+3)+\frac{\dd p}{\dd\ln a}
\label{eq:kappa}
\end{equation}
is the effective curvature parameter~\cite{Ismail:2024zbq}. For constant $p$ we have $\kappa=p(p+3)$, and the mode equation can be written as
\begin{equation}
v_k''+\left[k^2-\frac{\nu^2-\frac14}{\tau^2}\right]v_k=0,
\label{eq:inflationmode}
\end{equation}
with
\begin{equation}
\nu^2=
\begin{cases}
\dfrac94+\kappa=\left(\dfrac32+p\right)^2,
& \xi=0,\\[8pt]
\dfrac94-12\xi, & K=1.
\end{cases}
\label{eq:nu}
\end{equation}
We take $p>-3/2$ in the kinetic case and $\xi<3/16$ in the curvature case, choosing $\nu>0$. The normalized Bunch--Davies solution~\cite{Bunch:1978yq} is then given by
\begin{equation}
v_k^{\rm in}(\tau)
=\frac{\sqrt{-\pi\tau}}{2}\,
e^{i\pi(2\nu+1)/4}H^{(1)}_\nu(-k\tau),
\label{eq:inmode}
\end{equation}
where $H^{(1)}_\nu$ is the Hankel function of the first kind and the mode approaches $e^{-ik\tau}/\sqrt{2k}$ as $-k\tau\to\infty$.

Outside the horizon, $-k\tau\ll1$, the dominant mode satisfies
\begin{equation}
\eta_k\propto a^{\,\nu-3/2-p},
\label{eq:superhorizon}
\end{equation}
with $p=0$ in the curvature case. For the kinetic coupling, $\nu=3/2+p$, and the field approaches a constant outside the horizon. The kinetic history then sets the frozen spectrum without any subsequent superhorizon growth. A positive curvature coupling, $0<\xi<3/16$, instead damps both the fluctuations and the homogeneous displacement, which obey the same long-wavelength equation~\cite{Alonso-Alvarez:2018tus,Alonso-Alvarez:2019ixv}.

Across $\tau_e$ the field $\eta_k$ and its conjugate momentum $a^2K^2\eta_k'$ are continuous. However, the jump in $K'$ makes $v_k'$ discontinuous, and the matching data are given by
\begin{align}
V_k&\equiv v_k(\tau_e)=v_k^{\rm in}(\tau_e),
\nonumber\\
W_k&\equiv v_k'(\tau_e^+)
=v_k^{{\rm in}\prime}(\tau_e)-p\HI V_k.
\label{eq:jump}
\end{align}
For the curvature coupling, $p=0$ and $v_k'$ is continuous.

In radiation domination $K=1$ and $R=0$, and with the constant potential mass retained the mode equation $v_k''+(k^2+a^2m_\eta^2)v_k=0$ becomes the Weber equation in $y\equiv\tau+2/\HI$, where $a=\HI y$. In terms of $x\equiv k/\sqrt{m_\eta\HI}$, the normalized
positive-frequency out mode is given by
\begin{equation}
v_k^{\rm out}(y)
=\frac{e^{-\pi x^2/8}}{(2m_\eta\HI)^{1/4}}\,
D_{-(1+ix^2)/2}\!
\left(e^{i\pi/4}\sqrt{2m_\eta\HI}\,y\right),
\label{eq:outmode}
\end{equation}
where $D_a$ is the parabolic cylinder function. At late times $v_k^{\rm out}\simeq(2\omega_k)^{-1/2}\exp(-i\int^\tau\omega_k\,\dd\tau)$ up to a constant phase, with $\omega_k^2=k^2+a^2m_\eta^2$. For nonrelativistic modes we have $|\omega_k'/\omega_k^2|\simeq H/m_\eta$, and once $H\ll m_\eta$ the particle interpretation is adiabatic even for wavelengths larger than the Hubble radius.

After the transition, $v_k=\alpha_kv_k^{\rm out}+\beta_kv_k^{{\rm out}*}$, where
\begin{equation}
\begin{aligned}
\alpha_k&=i\left[
v_k^{{\rm out}*}W_k-v_k^{{\rm out}\prime*}V_k
\right]_{\tau_e},
\\
\beta_k&=-i\left[
v_k^{\rm out}W_k-v_k^{{\rm out}\prime}V_k
\right]_{\tau_e}.
\label{eq:Bogoliubov}
\end{aligned}
\end{equation}
The coefficients satisfy $|\alpha_k|^2-|\beta_k|^2=1$ and give the late-time occupation number $n_k=|\beta_k|^2$, which is exact for the specified background and kinetic history, with the potential mass neglected during inflation, and reduces to the minimally coupled calculation of
Ref.~\cite{Herring:2019hbe} at $p=\xi=0$.

If we also neglect the mass after inflation, the out modes are plane waves, and Eqs.~\eqref{eq:inmode} and~\eqref{eq:jump} give
\begin{align}
n_k &=\frac{|W_k|^2+k^2|V_k|^2}{2k}-\frac12
\nonumber\\
&=\frac{\pi}{8q}\Big[\big|qH_\nu'(q)+\big(p+\tfrac12\big)H_\nu(q)\big|^2
+q^2|H_\nu(q)|^2\Big]-\frac12,
\label{eq:nmassless}
\end{align}
where $q\equiv k/(a_e\HI)$, $H_\nu\equiv H_\nu^{(1)}$, and $p=0$ for the curvature coupling. For the canonical minimally coupled field ($\nu=3/2$) this is exactly $n_k=1/(4q^4)$, and at conformal coupling ($K=1$, $\xi=1/6$, $\nu=1/2$) it vanishes identically, as expected. However, the massive occupation number also depends on the relative phase of $V_k$ and $W_k$ and cannot be inferred from the massless result alone.

\subsection{Infrared spectrum}
\label{sec:IR}
For $q\ll1$, the field spectrum at the end of inflation, where $a_eK_e=1$, is
\begin{align}
\mathcal P_{\eta,e}(k)
&\equiv\frac{k^3}{2\pi^2}|V_k|^2
\simeq\mathcal C_\nu
\left(\frac{\HI}{2\pi}\right)^2q^{3-2\nu},
\nonumber\\
\mathcal C_\nu
&\equiv
\left[\frac{2^{\nu-3/2}\Gamma(\nu)}
{\Gamma(3/2)}\right]^2,
\label{eq:Pe}
\end{align}
with $\mathcal C_{3/2}=1$. We define the spectral tilt of the axion spectrum by
\begin{equation}
n_\eta-1\equiv\frac{\dd\ln\mathcal P_{\eta,e}}{\dd\ln k}=3-2\nu=
\begin{cases}
-2p, & \xi=0,\\[2pt]
3-\sqrt{9-48\xi}, & K=1.
\end{cases}
\label{eq:tilt}
\end{equation}
A positive $p$ gives a red spectrum, as earlier modes leave the horizon with a smaller $K$, and for the curvature coupling $n_\eta-1\simeq8\xi$ at small $\xi$.

To leading order in $q$ we find $W_k\simeq d\HI V_k$, where $d\equiv\nu-\frac12-p$. As long as gradients and the mass are negligible, $v_k''\simeq0$ in radiation domination, and $v_k=V_k+W_k(\tau-\tau_e)$. With $a=1+\HI(\tau-\tau_e)$, the axion mode $\eta_k=v_k/a$ then evolves as
\begin{equation}
\eta_k(a)\simeq
\left[d+\frac{1-d}{a}\right]\eta_{k,e},
\label{eq:IRtransfer}
\end{equation}
where $\eta_{k,e}=V_k$ is its value at the end of inflation, so $d\eta_{k,e}$ is the amplitude of the constant mode. The kinetic coupling has $d=1$, whereas the curvature coupling has $d=\nu-\frac12$, which exceeds unity for $\xi<0$, lies between zero and unity for $0<\xi<1/6$, and becomes negative for $1/6<\xi<3/16$.

\begin{figure}[t!]
\centering
\includegraphics[width=\columnwidth]{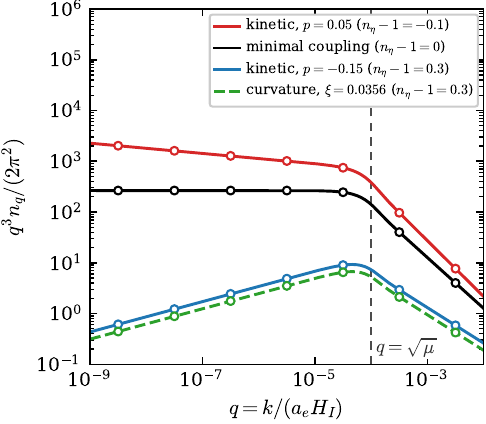}
\caption{
Late-time occupation spectra for $\mu\equiv m_\eta/\HI=10^{-8}$ and four choices of $p$ and $\xi$. Lines show Eq.~\eqref{eq:massiveIR}, and circles show the full matching, Eq.~\eqref{eq:Bogoliubov}. The turnover lies near $q=\sqrt\mu$.
}
\label{fig:spectra}
\end{figure}

With $\mu\equiv m_\eta/\HI$ and $n_q\equiv n_{k=qa_e\HI}$, the leading occupation spectrum for $q,\sqrt\mu\ll1$ and $|d|\gg\max(q,\sqrt\mu)$ is
\begin{equation}
\frac{q^3n_q}{2\pi^2}
\simeq
\frac{\cRD}{\sqrt\mu}\,
\frac{d^2\mathcal P_{\eta,e}(q)}{\HI^2}\,
\Ftr\!\left(\frac{q}{\sqrt\mu}\right),
\label{eq:massiveIR}
\end{equation}
where $\cRD\equiv\pi/[2\Gamma^2(3/4)]\simeq1.046$ and
\begin{equation}
\Ftr(x)=
\frac{\Gamma^2(3/4)e^{-\pi x^2/4}}
{\left|\Gamma\!\left(\frac34+\frac{ix^2}{4}\right)\right|^2}.
\label{eq:transfer}
\end{equation}
The left-hand side is the comoving number per logarithmic momentum interval in units of $(a_e\HI)^3$. The factor $d^2$ accounts for the transfer to the constant radiation era mode, and $\Ftr$ describes the subsequent massive evolution (App.~\ref{app:IR}). Note that the same $\cRD$ fixes the misalignment normalization, as we show in Sec.~\ref{sec:abundance}.

Using $\Ftr(0)=1$ and $\Ftr(x\gg1)\simeq\Gamma^2(3/4)/(\pi x)$, we find
\begin{equation}
n_q\propto
\begin{cases}
q^{-2\nu}, & q\ll\sqrt\mu,\\
q^{-2\nu-1}, & \sqrt\mu\ll q\ll1.
\end{cases}
\label{eq:IRslopes}
\end{equation}
The turnover occurs near the horizon scale at the onset of oscillations. Defining this onset by $H=m_\eta$ gives $a_{\rm osc}=\mu^{-1/2}$ and $k_{\rm osc}=a_e\sqrt{m_\eta\HI}$, or $q_{\rm osc}=\sqrt\mu$. Longer modes begin to oscillate outside the horizon and evolve locally like a misalignment field. Shorter modes instead enter the horizon while relativistic and redshift as radiation until they become nonrelativistic, which gives the extra factor $1/x$.

Setting $p=\xi=0$, we recover from Eq.~\eqref{eq:massiveIR} the minimally coupled spectrum of Ref.~\cite{Herring:2019hbe}, including its normalization. To this order, couplings with the same tilt give the same spectral shape and differ only by $d^2$. For example, $n_\eta-1=0.3$ corresponds to $p=-0.15$ in the kinetic case and to $\xi\simeq0.0356$ in the curvature case, with transfer coefficients $d=1$ and $d=0.85$, respectively, and the two normalizations differ by $d^2\simeq0.72$. In Fig.~\ref{fig:spectra} we compare Eq.~\eqref{eq:massiveIR} with the full matching.

Note that at conformal coupling $d=0$ and Eq.~\eqref{eq:massiveIR} does not apply. In this case the full matching gives $n_q\propto q^{-1}$ in the deep infrared~\cite{Herring:2019hbe}, as the nonzero potential mass breaks conformal invariance and allows production.

\subsection{Infrared cutoff}
\label{sec:IRcutoff}
We separate fluctuations within the present Hubble patch from longer modes, which contribute to its approximately homogeneous axion displacement. Following Refs.~\cite{Herring:2019hbe,Garcia:2022vwm}, we adopt a sharp split at $k_{\rm IR}=a_0H_0$, giving
\begin{equation}
q_{\rm IR}\equiv\frac{k_{\rm IR}}{a_e\HI}
=\frac{a_0H_0}{a_e\HI},
\label{eq:IRcutoff}
\end{equation}
where $a_0$ and $H_0$ are the present scale factor and Hubble rate. 

For instantaneous reheating followed by entropy conservation, we find
\begin{align}
q_{\rm IR}
&=\frac{H_0}{\HI}\frac{T_{\rm RH}}{T_0}
\left(\frac{g_{*s,e}}{g_{*s,0}}\right)^{1/3},
\nonumber\\
T_{\rm RH}
&=\left(\frac{90}{\pi^2g_{*,e}}\right)^{1/4}
\sqrt{\Mpl\HI},
\label{eq:IRthermal}
\end{align}
where $T_{\rm RH}$ is the temperature immediately after reheating, $T_0$ is the present photon temperature, and $g_*$ and $g_{*s}$ count the effective energy and entropy degrees of freedom. We assume that inflation began early enough for the lowest included mode to start well inside the horizon and in its adiabatic regime, $k_{\rm IR}/(a_i\HI)\gg\max(1,\sqrt{|\nu^2-1/4|})$. Longer inflation only adds modes to the patch mean and leaves the split unchanged.

Integrating Eq.~\eqref{eq:Pe} over $q_{\rm IR}<q<q_{\rm max}\ll1$, we obtain the variance at the end of inflation,
\begin{equation}
\langle\delta\eta^2\rangle_e
\simeq
\mathcal C_\nu\left(\frac{\HI}{2\pi}\right)^2
\begin{cases}
\dfrac{q_{\rm max}^{3-2\nu}-q_{\rm IR}^{3-2\nu}}
{3-2\nu}, & \nu\neq\frac32,\\[8pt]
\ln(q_{\rm max}/q_{\rm IR}), & \nu=\frac32.
\end{cases}
\label{eq:IRvariance}
\end{equation}
In the scale-invariant case with $\nu=3/2$, the variance measures the logarithmic range of modes included in the patch fluctuations. In the effectively massless limit, additional inflation before the patch scale left the horizon increases the variance of the patch mean~\cite{Vilenkin:1982wt, Linde:1982uu, Starobinsky:1982ee}.

The longest included mode exits the horizon $N_{\rm IR}=\ln(1/q_{\rm IR})$ $e$-folds before the end of inflation, and neglecting the potential mass requires its accumulated effect on the retained modes to remain small. For the constant-$p$ kinetic model, one can show that a sufficient condition is $\int\dd N\,m_\eta^2/[(3+2p)K^2\HI^2]\ll1$, with the integral taken over these $N_{\rm IR}$ $e$-folds.

For fixed $d\neq0$, the massive number integral has the same infrared behavior as the variance, i.e., logarithmic at $\nu=3/2$, power-law sensitive to $q_{\rm IR}$ for a red spectrum, and convergent as $q_{\rm IR}\to0$ for a blue one. Note that modes below $k_{\rm IR}$ also oscillate once $H<m_\eta$, without reentering the horizon, and contribute through the local displacement used for misalignment. We use the same split in Sec.~\ref{sec:abundance}.

\subsection{Ultraviolet sensitivity}
\label{sec:UVtail}
The sharp transition produces high-frequency tails, and for $q\gg1$ the massless result~\eqref{eq:nmassless} becomes
\begin{equation}
n_k\simeq
\begin{cases}
\dfrac{p^2}{4q^2}, & \xi=0,\ p\neq0,\\[8pt]
\dfrac{(1-6\xi)^2}{4q^4}, & K=1.
\end{cases}
\label{eq:UVoccupation}
\end{equation}
The $q^{-2}$ tail comes from the jump in $K'$ and makes the comoving number diverge linearly. The $q^{-4}$ tail comes from the jump in $a''$, and the number then converges, but the relativistic energy diverges logarithmically unless $\xi=1/6$. These tails are artifacts of the sharp transition and are suppressed for $k/a_e\gg\Delta t^{-1}$ when the exit lasts a time $\Delta t$.

For the quadratic exponential, an exit at $\phi_e=0$ gives $K'_e=0$ and removes the $q^{-2}$ tail, although a nonzero $K''_e$ can still change the coefficient of the $q^{-4}$ tail. As a check, we integrated Eq.~\eqref{eq:mode} through a smooth exit over the last $\Delta N\leq2$ $e$-folds, with $K=1$ and $K'_e=K''_e=0$ at the transition, and found that for $p=-0.15$ and $\mu=10^{-8}$ the $q^{-2}$ tail is suppressed at high momentum, while modes with $q\leq10^{-2}$ change by less than $0.1\%$.

Above $q\sim1$ the result depends on how reheating proceeds, and for the kinetic coupling the sharp transition makes the number diverge. The abundance is then reliable only if these modes contribute little, which is the case here, as $q^3n_q$ scales as $q^{3-2\nu}$ below the turnover and as $q^{2-2\nu}$ above it and, for $1<\nu<3/2$, peaks near $q\sim\sqrt\mu$, far below $q\sim1$. Within the instantaneous transition one can follow the spectrum up to $q\sim1$, as the mass plays no role at the matching for $q\gg\sqrt\mu$.\footnote{Eqs.~\eqref{eq:nmassless} and~\eqref{eq:massiveIR} agree at $q=10^{-2}$ to $4\times10^{-5}$, and we integrate Eq.~\eqref{eq:massiveIR} up to $q_{\rm max}=10^{-2}$. Extending the integral to $q=1$, as in Ref.~\cite{Herring:2019hbe}, adds $1.1\%$ to the abundance at $n_\eta-1=0.3$ ($1.3\%$ for the curvature coupling) and $10\%$ at $n_\eta-1=0.5$.}

\section{Relic abundance}
\label{sec:abundance}
\label{sec:split}
In a comoving patch we write
\begin{equation}
\eta=\bar\eta(t)+\delta\eta,\qquad\eta_c=f_\eta\theta_c ,
\label{eq:split}
\end{equation}
where the mean $\bar\eta$ comes from modes longer than the patch and $\delta\eta$ from shorter ones. As discussed in Sec.~\ref{sec:IRcutoff}, the patch is the present Hubble volume. Before the mass term becomes important, the radiation era transfer~\eqref{eq:IRtransfer} leaves $\bar\eta$ as a constant mode, with amplitude $\eta_c$, and a mode decaying as $1/a$, and we assume that the decaying mode is negligible by the time the field begins to oscillate.

Fluctuations in the band $q_{\rm IR}\leq q\leq q_{\rm max}$ have comoving number density $a^3n_\eta^{\rm fluc}=\HI^3\Iprod$, with
\begin{equation}
\Iprod\equiv\frac{1}{2\pi^2}\int_{q_{\rm IR}}^{q_{\rm max}}\dd q\,q^2n_q .
\label{eq:I}
\end{equation}
In all cases studied below, the modes that dominate the abundance become nonrelativistic well before matter--radiation equality, and those that are still relativistic at that time carry a negligible share of the energy. We can then take $\rho_\eta^{\rm fluc}\simeq m_\eta n_\eta^{\rm fluc}$.

After reheating the entropy density is $s_e=(2\pi^2/45)g_{*s,e}T_{\rm RH}^3$, with $T_{\rm RH}$ from Eq.~\eqref{eq:IRthermal}. Once production has ended, the yield $n_\eta^{\rm fluc}/s=\HI^3\Iprod/s_e$ stays constant, provided the axion is stable and decoupled and entropy is conserved, and the fluctuation abundance is given by
\begin{equation}
\Omega^{\rm fluc}_\eta h^2=2.744\times10^{8}\left(\frac{m_\eta}{\mathrm{GeV}}\right)\frac{\HI^3\Iprod}{s_e}.
\label{eq:Omega}
\end{equation}

For misalignment we use the regular homogeneous solution, treating the relativistic degrees of freedom as constant through the onset of oscillations. In radiation domination, with the time origin chosen so that $a^2=2\HI t$, we have $H=1/(2t)$ and find
\begin{equation}
\frac{\bar\eta(t)}{\eta_c}=\Gamma(5/4)\left(\frac{m_\eta t}{2}\right)^{-1/4}J_{1/4}(m_\eta t),
\label{eq:misexact}
\end{equation}
which approaches unity as $m_\eta t\to0$. The other solution behaves as $t^{-1/2}\propto1/a$ at early times and is the decaying mode we have neglected. Averaging over oscillations at $m_\eta t\gg1$, we find $a^3\rho^{\rm mis}_\eta=\cRD\,\eta_c^2\,m_\eta^{1/2}\HI^{3/2}$, with $\cRD=4\Gamma^2(5/4)/\pi=\pi/[2\Gamma^2(3/4)]$ as in Eq.~\eqref{eq:massiveIR}. The misalignment abundance can then be written as
\begin{equation}
\Omega^{\rm mis}_\eta h^2=2.744\times10^{8}\,
\frac{\cRD\,\eta_c^2\,m_\eta^{1/2}\HI^{3/2}}{s_e\,\mathrm{GeV}} .
\label{eq:Omis}
\end{equation}
As $s_e\propto(\Mpl\HI)^{3/2}$, the displacement that gives the observed abundance does not depend on $\HI$,
\begin{equation}
\eta_{\rm DM}=1.08\times10^{10}\GeV\left(\frac{10\,\mathrm{MeV}}{m_\eta}\right)^{1/4}\left(\frac{g_{*,e}}{106.75}\right)^{1/8},
\label{eq:etaDM}
\end{equation}
for $g_{*s,e}=g_{*,e}$.

In terms of $\eta_{\rm DM}$, Eq.~\eqref{eq:Omis} reads $\Omega^{\rm mis}_\eta h^2=0.12\,\eta_c^2/\eta_{\rm DM}^2$. Comparing Eqs.~\eqref{eq:Omega} and~\eqref{eq:Omis}, we find that the fluctuations give the same abundance as a homogeneous displacement $\sqrt Q$, with
\begin{equation}
Q\equiv\frac{\HI^2\sqrt\mu}{\cRD}\,\Iprod ,
\label{eq:Qdef}
\end{equation}
so that $\Omega^{\rm fluc}_\eta h^2=0.12\,Q/\eta_{\rm DM}^2$ and
\begin{equation}
\frac{\Omega^{\rm fluc}_\eta}{\Omega^{\rm mis}_\eta}=\frac{Q}{f_\eta^2\theta_c^2}.
\label{eq:dominance}
\end{equation}
Fluctuations alone then give the observed abundance at the mass $m_*$ where $Q=\eta_{\rm DM}^2$. Using Eq.~\eqref{eq:massiveIR}, $Q$ can be written as
\begin{equation}
Q\simeq d^2\int_{q_{\rm IR}}^{q_{\rm max}}\dd\ln q\;\mathcal P_{\eta,e}(q)\,\Ftr(x),\qquad x\equiv q/\sqrt\mu .
\label{eq:Qintegral}
\end{equation}

\begin{figure}[t!]
\centering
\includegraphics[width=\columnwidth]{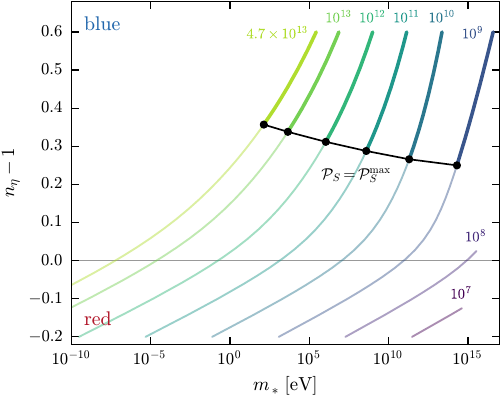}
\caption{The axion mass $m_*$ at which the fluctuations alone give $\Omega_\eta h^2=0.12$, as a function of the tilt, for constant kinetic slope. Curves are labeled by $\HI$ in GeV, and $4.7\times10^{13}\GeV$ is the tensor bound, $\HI<\pi\Mpl\sqrt{rA_s/2}$ with tensor-to-scalar ratio $r<0.036$~\cite{BICEP:2021xfz} and scalar amplitude $A_s = 2.1 \times 10^{-9}$. Thick segments satisfy the pivot isocurvature benchmark~\eqref{eq:isobound} with zero mean, thin segments exceed it, and black dots mark the threshold $\PS=\PS^{\max}$. The smallest allowed mass is $4.5\,$keV at $\HI=10^{13}\GeV$ and $0.14\,$keV at the tensor bound.}
\label{fig:mstar}
\end{figure}

The modes that set $Q$ depend on the tilt. For a red spectrum the longest modes dominate, and $Q$ depends on $q_{\rm IR}$. For a flat spectrum every logarithmic interval below the turnover contributes equally, and $Q\propto\ln(\sqrt\mu/q_{\rm IR})$. For a blue spectrum, $1<\nu<3/2$, the contribution per logarithmic interval peaks near $q\sim\sqrt\mu$ and the integral converges at both ends. In this case the cutoffs contribute little for $q_{\rm IR}\ll\sqrt\mu\ll q_{\rm max}\ll1$, and extending the integral to $0<x<\infty$ gives
\begin{equation}
Q\simeq d^2\mathcal C_\nu\left(\frac{\HI}{2\pi}\right)^2\mathcal J\,\mu^{(n_\eta-1)/2},
\label{eq:Qblue}
\end{equation}
with $\mathcal J\equiv\int_0^\infty\dd x\,x^{2-2\nu}\Ftr(x)$. For fixed $g_*$, the abundance then scales as
\begin{equation}
\Omega^{\rm fluc}_\eta\propto\HI^{\,2-(n_\eta-1)/2}\,m_\eta^{\,n_\eta/2}.
\label{eq:scaling}
\end{equation}
As $\mu\ll1$, a bluer spectrum gives less abundance at a given mass and requires a heavier axion.

Ref.~\cite{Turner:1987vd} estimated this abundance for a light scalar with arbitrary $\xi$ by following modes through horizon exit and reentry, and found it dominated by modes near $k\simeq k_{\rm osc}$. In their estimate a mode still outside the horizon at the onset of oscillations carries only its gradient energy at that time, which suppresses it by $(k/k_{\rm osc})^2$. Its potential energy $m_\eta^2|\eta_k|^2/2$ is larger by a factor of order $(k_{\rm osc}/k)^2$, and these modes evolve like a local misalignment field, $\Ftr\simeq1$. Applied to the QCD axion with $f_\eta=\HI$, which maximizes their abundance, the estimate makes the fluctuations important for $\xi\lesssim-0.033$. Such couplings give red spectra, which exceed our pivot benchmark by many orders of magnitude if the fluctuations account for all of the dark matter (Sec.~\ref{sec:iso}).\footnote{Ref.~\cite{Turner:1987vd} suppresses the long modes that dominate $Q$ for flat and red spectra, and so misses most of the abundance. For a flat spectrum these modes contribute $1/2$ to the bracket in its Eq.~(3.10), instead of the logarithm of the range of modes below $k_{\rm osc}$. It also ignores isocurvature, which now excludes the red spectra that give a large abundance there. Its axion estimate takes $f_\eta=\HI$ with instantaneous reheating, so that $T_{\rm RH}\gg f_\eta$ and a thermalized PQ sector would restore the symmetry, which closes the QCD window (Sec.~\ref{sec:QCD}).}

We take a low and a high inflationary scale, $\HI=10^{10}$ and $10^{13}\GeV$, with $g_{*,e}=g_{*s,e}=106.75$, $g_{*s,0}=3.91$, $T_0=2.7255\,$K, $h=0.674$, and $q_{\rm max}=10^{-2}$. The CMB pivot $k_*=0.05\Mpc$ then leaves the horizon $N_*\equiv\ln(a_e\HI/k_*)=51.7$ and $55.2$ $e$-folds before the end of inflation, from Eq.~\eqref{eq:IRthermal} with $a_0H_0$ replaced by $k_*$.\footnote{At $m_*$ for $0\leq n_\eta-1\leq0.5$ at both scales, lowering $q_{\rm max}$ from $10^{-2}$ to $10^{-3}$ changes the abundance by less than $5\%$, except for $n_\eta-1=0.4$ and $0.5$ at the lower scale, where the reduced cutoff lies close to the turnover and the abundance drops by $12\%$ and $25\%$.}

The mass $m_*$ rises steeply with the tilt and, from Eq.~\eqref{eq:scaling}, falls with the inflationary scale as $m_*\propto\HI^{-(5-n_\eta)/n_\eta}$ (see Fig.~\ref{fig:mstar}). For example, for $n_\eta-1=0.3$ we find $m_*=384\GeV$ at $\HI=10^{10}\GeV$, where $\sqrt Q=\eta_{\rm DM}=7.7\times10^{8}\GeV$, and $m_*=1.1\,$keV at $\HI=10^{13}\GeV$. The curvature coupling with the same tilt has $d^2\simeq0.72$, which lowers its abundance at fixed mass by this factor and raises its $m_*$ by about $1.65$.

These masses require long-lived, weakly coupled ALPs. A photon coupling $g_{\eta\gamma\gamma}=\alpha_{\rm em}/(2\pi f_\eta)$, for example, with $f_\eta=10^{12}\GeV$ gives a photon-decay lifetime of only $1.7\,\mathrm{s}$ at $m_\eta=384\GeV$. An exact $\eta\to-\eta$ symmetry can forbid single-axion decays, but it does not by itself suppress thermal pair production. The QCD axion, whose mass is fixed by $f_\eta$ and depends on temperature, does not lie on these curves, and we treat it separately in Sec.~\ref{sec:QCD}. For $n_\eta-1\geq0.1$ at both scales the field starts to oscillate above $10^3\GeV$, where $g_*$ is constant and Eq.~\eqref{eq:misexact} holds.

\section{Isocurvature}
\label{sec:iso}
Once the axion behaves as pressureless matter, its entropy perturbation relative to the radiation is, at linear order, given by
\begin{equation}
\mathcal{S}_{\eta}=3(\zeta_\eta-\zeta_\gamma)=\frac{\delta\rho_\eta}{\bar\rho_\eta}-\frac34\frac{\delta\rho_\gamma}{\bar\rho_\gamma},
\label{eq:Sdef}
\end{equation}
where $\zeta_i$ is the curvature perturbation on uniform-density slices of species $i$. Adiabaticity is a property of the perturbations, not of the mechanism that set the mean density~\cite{Weinberg:2003sw,Allali:2025pja}, and an axion made of vacuum fluctuations need not be adiabatic. Inflationary axion isocurvature was studied in Refs.~\cite{Axenides:1983hj,Seckel:1985tj}. However, since the density is quadratic in the field, a vanishing mean removes the linear contribution but leaves a quadratic one. At zero mean the leading density perturbation is therefore quadratic in the field, and by linear order we mean linear in $\delta\rho_\eta$, not in $\delta\eta$.

Modes that are still outside the horizon when the oscillations begin, $k_{\rm IR}<k\ll k_{\rm osc}$, are frozen at that time and shift the local misalignment displacement from $\eta_c$ to $\eta_c+X$. Here $X$ is the long-wavelength part of $\delta\eta$ after the radiation era transfer~\eqref{eq:IRtransfer}, in which each mode enters with its constant amplitude $d\,\eta_{\bm k,e}$, and $X$ is Gaussian with
\begin{equation}
\mathcal P_X(k)=d^2\,\mathcal P_{\eta,e}(k)=d^2\mathcal C_\nu\left(\frac{\HI}{2\pi}\right)^2q^{3-2\nu}.
\label{eq:PX}
\end{equation}
On scales $k\ll k_{\rm osc}$ the long-mode density on a uniform-radiation slice is then proportional to $(\eta_c+X)^2$, and including the short modes in the mean density gives
\begin{equation}
\frac{\delta\rho_\eta}{\bar\rho_\eta}\simeq\frac{2\eta_cX+X^2-\langle X^2\rangle}{\eta_c^2+Q} .
\label{eq:quadraticdensity}
\end{equation}
The numerator omits the short modes, which we estimate below, and $\langle X^2 \rangle < Q$ because $X$ contains only the long modes. If the rest of the dark matter is adiabatic, the total cold dark matter entropy perturbation is $\mathcal{S}=(\Omega_\eta/\Omega_c) \mathcal{S}_\eta$, with $\Omega_c$ the cold dark matter density, and its power spectrum follows from Wick's theorem,
\begin{equation}
\PS(k)\simeq\left(\frac{\Omega_\eta}{\Omega_c}\right)^2\frac{4\eta_c^2\,\mathcal P_X(k)+\mathcal P_{X^2}(k)}{(\eta_c^2+Q)^2},
\label{eq:Wick}
\end{equation}
with
\begin{equation}
\mathcal P_{X^2}(k)=\frac{k^3}{\pi^2}\int\frac{\dd^3\ell}{(2\pi)^3}\,P_X(\ell)\,P_X(|\bm k-\bm\ell|)
\label{eq:PX2}
\end{equation}
and $P_X=2\pi^2\mathcal P_X/k^3$. The first term describes the interference between the mean and the fluctuations. The second term survives at $\eta_c=0$, where $\mathcal{S}_\eta$ is, in the long-mode approximation, a shifted and rescaled $\chi^2$ field~\cite{Linde:1996gt, Peebles:1998nu, Kawasaki:2008sn,Langlois:2008vk}.

Modes with $q\gtrsim\sqrt\mu$ require the momentum-dependent kernel of App.~\ref{app:density}. On scales $k\ll k_{\rm osc}$ they give white noise, $\PS\propto k^3$, which is subdominant at small $k$ for $\nu>3/4$. Note that white noise dominates only when the short modes dominate the convolution, as for superheavy dark matter~\cite{Chung:2004nh,Racco:2024aac,Verner:2024agh}.

For $3/4<\nu<3/2$ and $q_{\rm IR}\ll q_*\ll\sqrt\mu$, with $q_*=e^{-N_*}$, the convolution converges at both ends. We nevertheless keep the infrared cutoff, whose effect remains appreciable near scale invariance. Modes longer than the present horizon belong to $\eta_c$, and both momenta in Eq.~\eqref{eq:PX2} must satisfy
$|\bm\ell|,|\bm k-\bm\ell|>k_{\rm IR}$, which gives
\begin{equation}
\PS(k_*)\simeq\left(\frac{\Omega_\eta}{\Omega_c}\right)^2\frac{4\eta_c^2\,\mathcal P_X(k_*)+\mathcal D_c\,\mathcal P_X^2(k_*)}{(\eta_c^2+Q)^2},
\label{eq:isoanalytic}
\end{equation}
where $\mathcal D_c(\nu;q_{\rm IR}/q_*)$ is given in App.~\ref{app:density} and corrections from the short modes are suppressed by $(q_*/\sqrt\mu)^{4\nu-3}$. Without the cutoff,
\begin{equation}
\mathcal D(\nu)=\frac{\sqrt\pi}{2}\,\frac{\Gamma^2\!\left(\frac32-\nu\right)\Gamma\!\left(2\nu-\frac32\right)}{\Gamma^2(\nu)\,\Gamma(3-2\nu)},
\qquad\frac34<\nu<\frac32 .
\label{eq:Dnu}
\end{equation}

\begin{figure*}[ht!]
\centering
\includegraphics[width=\textwidth]{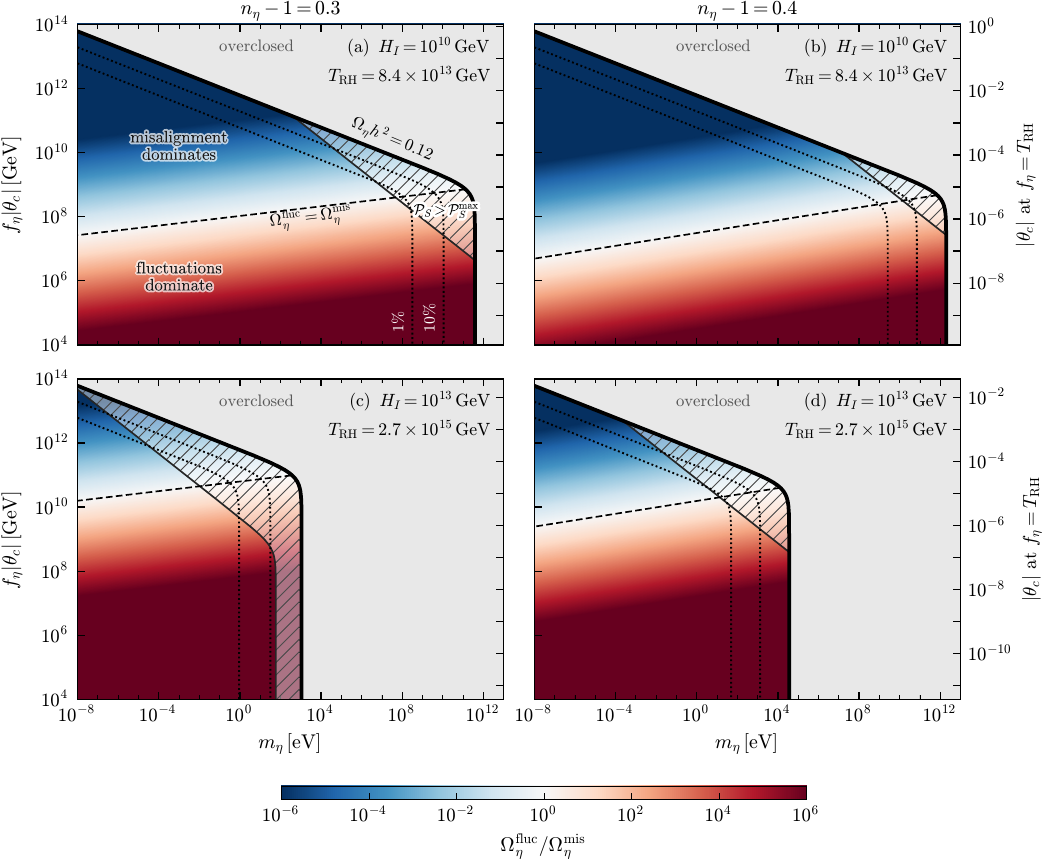}
\caption{Fluctuation to misalignment abundance ratio for the kinetic coupling at $\HI=10^{10}\GeV$ (top) and $10^{13}\GeV$ (bottom), for $n_\eta-1=0.3$ (left) and $0.4$ (right). The color shows $\Omega^{\rm fluc}_\eta/\Omega^{\rm mis}_\eta=Q/(f_\eta\theta_c)^2$, and the dashed line marks equal contributions. The thick black line corresponds to $\Omega_\eta h^2=0.12$ and above it, in gray, the axion overcloses the universe, and the line drops vertically at $m_*$. Thin dotted lines mark $\Omega_\eta=0.1\,\Omega_c$ and $0.01\,\Omega_c$, labeled $10\%$ and $1\%$ in panel (a). Hatched regions exceed the pivot benchmark~\eqref{eq:isobound}, with $\PS$ from Eq.~\eqref{eq:Wick} and $\Omega_\eta/\Omega_c=(\eta_c^2+Q)/\eta_{\rm DM}^2$. The right axes give $|\theta_c|$ for $f_\eta=T_{\rm RH}$, the approximate thermal restoration boundary for $T_c\simeq f_\eta$, with $T_{\rm RH}$ listed in each panel.}
\label{fig:iso}
\end{figure*}

\begin{figure*}[ht!]
\centering
\includegraphics[width=\textwidth]{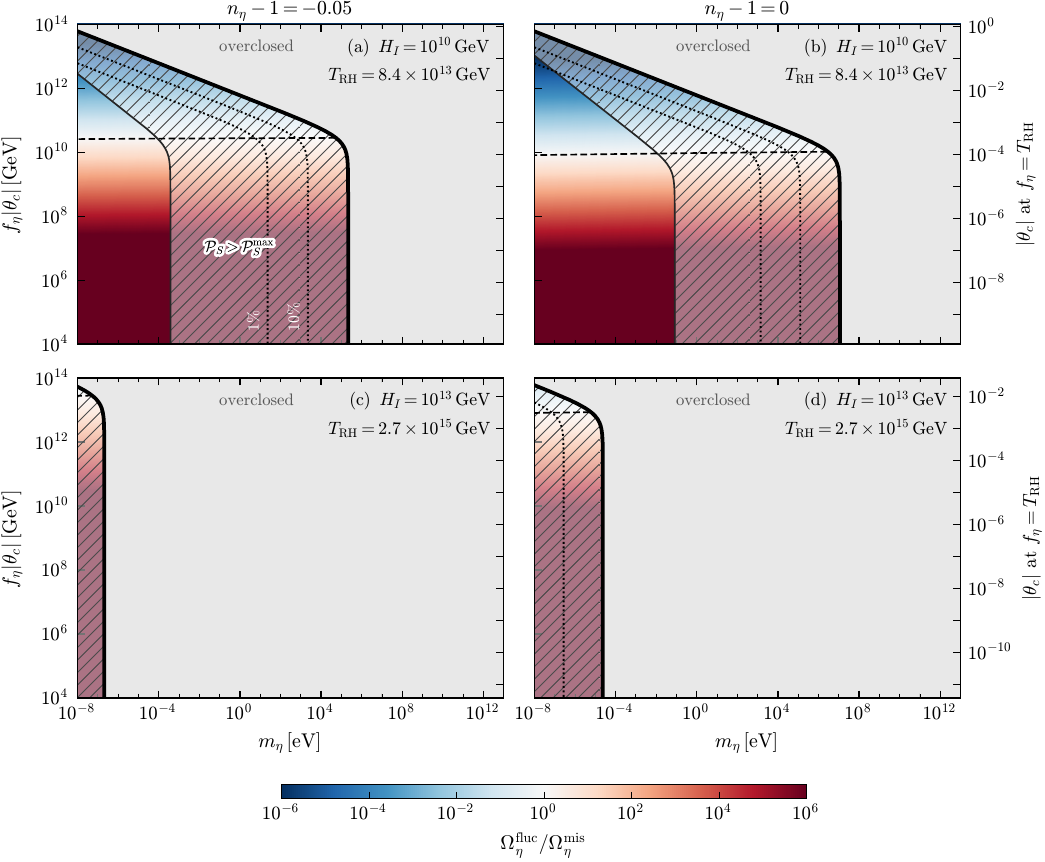}
\caption{Same as Fig.~\ref{fig:iso} for a red spectrum, $n_\eta-1=-0.05$ (left), and a flat one, $n_\eta-1=0$ (right). The fluctuations alone supply the dark matter at $m_*=2.2\times10^{5}\,$eV and $1.1\times10^{7}\,$eV for $\HI=10^{10}\GeV$, and at $2.1\times10^{-7}\,$eV and $2.4\times10^{-5}\,$eV for $\HI=10^{13}\GeV$. Along the whole dark matter line shown the isocurvature exceeds the pivot benchmark. Only a small, subdominant axion component at low mass satisfies it.}
\label{fig:iso_redflat}
\end{figure*}

If the axion makes up all of the dark matter and the mean vanishes,
\begin{equation}
\PS(k_*)\simeq\mathcal D_c\,\frac{\mathcal P_X^2(k_*)}{Q^2}.
\label{eq:PSzero}
\end{equation}
Up to $\mathcal D_c$, this is the square of $\mathcal P_X(k_*)/Q$, the fraction of $Q$ per logarithmic interval at the pivot. For $1<\nu<3/2$ and negligible endpoint corrections, Eq.~\eqref{eq:Qblue} turns it into $(\mathcal D_c/\mathcal J^2)(q_*/\sqrt\mu)^{6-4\nu}$, and the pivot power is set by the separation between the CMB and oscillation scales. Far above the cutoff the isocurvature spectrum is blue, with tilt $2(n_\eta-1)$.

As a benchmark, we use the Planck limit on uncorrelated, scale-invariant cold dark matter isocurvature at the pivot, $\beta_{\rm iso}<0.038$~\cite{Planck:2018jri}, which for $A_s=2.1\times10^{-9}$ gives\footnote{Analyses that include ACT data find limits slightly weaker than Planck alone at the pivot for blue spectra~\cite{Petretti:2026ayw}.}
\begin{equation}
\PS(k_*)<\PS^{\max}\equiv\frac{\beta_{\rm iso}}{1-\beta_{\rm iso}}\,A_s=8.3\times10^{-11}.
\label{eq:isobound}
\end{equation}
Fig.~\ref{fig:iso} shows the abundance ratio, the dark matter line, and the region that exceeds this benchmark for blue spectra.\footnote{For the blue non-Gaussian spectra found here this is a benchmark and not a likelihood bound, and a full analysis would also include the CMB constraints on non-Gaussian isocurvature~\cite{Hikage:2012be}.}

When the fluctuations make up all of the dark matter, the flat and mildly red spectra of Fig.~\ref{fig:iso_redflat} give $\PS(k_*)\sim10^{-2}$--$10^{-1}$, as their long-wavelength power is not suppressed. Blue spectra move $Q$ to small scales, and at $\HI=10^{10}\GeV$ we find that they satisfy the benchmark for
\begin{equation}
\begin{aligned}
n_\eta-1>0.27\quad&(\text{kinetic, constant }p),\\
\xi>0.032\quad&(\text{curvature}),
\label{eq:tc}
\end{aligned}
\end{equation}
with $m_\eta\gtrsim200\GeV$ and $310\GeV$, respectively. The curvature threshold corresponds to $n_\eta-1>0.26$, in agreement with $\xi\gtrsim0.03$ found in Ref.~\cite{Alonso-Alvarez:2019ixv}. Above the threshold the pivot power falls quickly, by five orders of magnitude at $n_\eta-1=0.4$, and for $n_{\eta} - 1 = 0.3$, with zero mean we find $\mathcal{P}_S(k_*) \simeq 4 \times 10^{-12}$, with the isocurvature power exceeding the adiabatic power $A_s$ only for $k \gtrsim 10^{3} \, \rm{Mpc}^{-1}$. As shown in Fig.~\ref{fig:mstar}, the threshold increases with the scale of inflation, from $n_\eta-1>0.26$ at $\HI=5\times10^{9}\GeV$ to $0.34$ at $\HI=10^{13}\GeV$ (where $m_{\eta} \gtrsim 4.5 \, \rm{keV}$) and $0.36$ at the tensor bound, $\HI=4.7\times10^{13}\GeV$~\cite{BICEP:2021xfz}. Over the same range the threshold mass decreases from $1.4\,$TeV to $0.14\,$keV. Below $5\times10^{9}\GeV$ the turnover approaches $q_{\rm max}$ and the fixed band becomes sensitive to the cutoff.

On smaller scales the isocurvature grows but is still allowed. For $n_{\eta} - 1 = 0.3$ at $H_I = 10^{10} \, \rm{GeV}$ we find $\mathcal{P}_S \simeq 3 \times 10^{-11}$ at $k = 1 \, \rm{Mpc}^{-1}$, where the Lyman-$\alpha$ forest gives the strongest small-scale limit~\cite{Buckley:2025zgh}, and $\mathcal{P}_S \simeq 8 \times 10^{-9}$ at $k = 10^{4} \, \rm{Mpc}^{-1}$. On these scales the $\mu$ distortion from cold dark matter isocurvature is suppressed by $(k_{\rm eq}/k)^2$~\cite{Buckley:2025zgh}, and for this spectrum we estimate the $\mu$-type distortion is at most of order $\lesssim 10^{-15}$, many orders of magnitude below the COBE/FIRAS bound.

Along the dark matter line, $\eta_c^2+Q=\eta_{\rm DM}^2$, a residual mean adds the linear term, which on the fluctuation-dominated branch limits how large the mean can be. In Fig.~\ref{fig:iso} the fluctuation-dominated end of the dark matter line satisfies the benchmark for $n_\eta-1=0.3$ only at $\HI=10^{10}\GeV$, and for $n_{\eta} - 1 = 0.4$ at both scales. We find that at $\HI=10^{10}\GeV$ the benchmark requires $\Omega^{\rm mis}_\eta/\Omega^{\rm fluc}_\eta<3.1\times10^{-5}$, $2.7\times10^{-3}$, and $0.52$ for $n_\eta-1=0.3$, $0.4$, and $0.5$, respectively. Larger means are excluded until misalignment dominates by a wide margin, $\Omega^{\rm mis}_\eta/\Omega^{\rm fluc}_\eta>6.7\times10^{5}$, $3.8\times10^{3}$, and $4.4$. For $\HI=10^{13}\GeV$ and $n_\eta-1=0.4$, where $m_*=36\,$keV, the corresponding bound is $8\times10^{-5}$. For $f_{\eta} = 10^{12} \, \rm{GeV}$ these bounds require $|\theta_c| \lesssim 4 \times 10^{-6}$ at $H_I = 10^{10} \, \rm{GeV}$ and $n_{\eta} - 1 = 0.3$, and $|\theta_c| \lesssim 4 \times 10^{-4}$ at $H_I = 10^{13} \, \rm{GeV}$ and $n_{\eta} - 1 = 0.4$.

In the kinetic realization such a small mean is an assumption about the initial conditions, unless the axion is heavy before the observable scales leave the horizon and relaxes to the same minimum it has at late times. In the curvature realization the small mean arises dynamically. If inflation lasted long enough before the patch scale left the horizon, $\xi>0$ erases the pre-inflationary displacement~\cite{Alonso-Alvarez:2018tus}, and leaves only the modes longer than the present horizon, $\langle\eta_c^2\rangle\simeq\mathcal P_X(k_{\rm IR})/(3-2\nu)$. For the curvature coupling with $n_\eta-1=0.3$ at $\HI=10^{10}\GeV$ this gives $\Omega^{\rm mis}_\eta/\Omega^{\rm fluc}_\eta\simeq4\times10^{-7}$, well below the bound.\footnote{For a finite duration $\Delta N$ before the patch scale exits, $1/(3-2\nu)$ becomes $[1-e^{-(3-2\nu)\Delta N}]/(3-2\nu)$, and any surviving initial displacement adds to $\eta_c$.}

In the opposite limit, $\eta_c^2\gg Q$, only the linear term in Eq.~\eqref{eq:isoanalytic} survives. Using $K_*=q_*^{\,p}$ for the kinetic coupling and Eq.~\eqref{eq:superhorizon} for the curvature coupling, we obtain
\begin{equation}
\PS(k_*)=\left(\frac{\Omega_\eta}{\Omega_c}\right)^2\left(\frac{2}{\theta_c}\right)^2d^2\,\mathcal C_\nu\,e^{-2(1-d)N_*}\left(\frac{\HI}{2\pi f_I}\right)^2,
\label{eq:response}
\end{equation}
with $f_I=K_*f_\eta$ from Eq.~\eqref{eq:fI}. For $d=\mathcal C_\nu=1$ and $f_I=f_\eta$ this is the standard result~\cite{Seckel:1985tj,Lyth:1989pb,Turner:1990uz,Hertzberg:2008wr}, and with $f_I$ and the transfer factor included it agrees with Refs.~\cite{Graham:2025iwx,Verner:2026cbn}. Beyond the harmonic regime, $2/\theta_c$ is replaced by $\partial\ln\Omega^{\rm mis}_\eta/\partial\theta_c$.

The kinetic coupling has $d=1$ and acts through $f_I$. For $p=-0.15$ at $\HI=10^{10}\GeV$, we have $K_*=e^{0.15N_*}\simeq2.3\times10^{3}$, and at fixed $\theta_c$ the isocurvature is smaller than in the canonical case by $\mathcal C_\nu/K_*^2$. This is the evolving radius mechanism~\cite{Linde:1991km,Kearney:2016vqw,Graham:2025iwx} written as a kinetic function. The curvature coupling has $K=1$ and acts instead through the factor $d\sqrt{\mathcal C_\nu}\,e^{-(1-d)N_*}$, which is $3.3\times10^{-4}$ for $n_\eta-1=0.3$ and $\HI=10^{10}\GeV$~\cite{Alonso-Alvarez:2019ixv}. However, it requires a larger initial displacement, by $1/[d\,e^{-(1-d)N_{\rm tot}}]$, which grows exponentially with the total number of $e$-folds $N_{\rm tot}$.

\subsection{The QCD axion}
\label{sec:QCD}
The QCD axion has $m_\eta\simeq5.7\,\mu{\rm eV}\,(10^{12}\GeV/f_\eta)$~\cite{GrillidiCortona:2015jxo}, and its temperature-dependent mass starts the oscillations at $T_{\rm osc}\sim0.2$--$1.5\GeV$ for the decay constants considered here~\cite{Borsanyi:2016ksw}. Equation~\eqref{eq:Qdef} then no longer applies, and we define $Q$ through Eq.~\eqref{eq:dominance} instead. Modes outside the horizon at $T_{\rm osc}$ oscillate with the zero mode and enter $Q$ directly. For shorter modes we use the constant mass transfer~\eqref{eq:transfer}, and the results below are estimates. Taking $T_{\rm osc}=1\GeV$, we find $k_*/k_{\rm osc}\simeq3\times10^{-9}$, independent of $\HI$.

Within this estimate, a single power law cannot give the full abundance while satisfying the pivot benchmark. Misalignment gives $\Omega^{\rm mis}_\eta\propto\theta_c^2f_\eta^{7/6}$~\cite{DiLuzio:2020wdo}, which implies $\eta_{\rm DM}\propto f_\eta^{5/12}$, and even at the edge of the harmonic regime, $\theta_{\rm DM}\equiv\eta_{\rm DM}/f_\eta\simeq1$, one needs $\eta_{\rm DM}\gtrsim10^{11}\GeV$. A flat spectrum, with $\sqrt Q\simeq0.8\,\HI$, reaches this value for $\HI\gtrsim10^{11}\GeV$, but then $\PS(k_*)\simeq3\times10^{-2}$. A blue spectrum satisfies the benchmark only for $n_\eta-1\gtrsim0.6$, and in this case $\sqrt Q\lesssim3\times10^{8}\GeV$ even at the tensor bound. The fluctuations then supply less than $10^{-5}$ of the dark matter, and a curvature coupling, with $d^2 < 1$, gives even less.

The running slope of the quadratic exponential (Sec.~\ref{sec:slopes}) improves this situation. At the tensor bound we find that the pivot benchmark allows $N_{\gamma} \lesssim 12.6$, corresponding to a pivot tilt of $0.70$ and a local tilt of $0.46$ near $k_{\rm osc}$, and the same estimate gives $\sqrt{Q} \simeq 2 \times 10^{11} \, \rm{GeV}$, comparable to the smallest $\eta_{\rm DM}$. Within our estimate the running case therefore remains marginal.

However, if we adopt $T_c\simeq f_\eta$ for the PQ restoration temperature, thermal non-restoration after instantaneous reheating requires $f_\eta\gtrsim T_{\rm RH}\simeq6\times10^{15}\GeV$ at the tensor bound. Within the misalignment estimate used here this raises $\eta_{\rm DM}$ to order $10^{13}\GeV$, and the fluctuations fall short also for the running slope. A different PQ thermal potential would require a separate analysis. The same is true for a slower reheating, which changes $N_*$, the entropy density $s_e$, and the transfer of the modes that start to oscillate before reheating ends, and which also lowers $T_{\rm RH}$ and with it the bound $f_{\eta} \gtrsim T_{\rm RH}$.

These conclusions assume that the axion stays light during inflation. If it is heavy while the CMB scales leave the horizon and becomes light later, the CMB scale isocurvature is removed and the later fluctuations can set the QCD axion abundance~\cite{Kitano:2023mra,Freese:2026xax}, a case that a single power law does not describe.

In the misalignment--dominated regime the couplings still help. Eq.~\eqref{eq:response} applies with the QCD relation between $\theta_c$ and $f_\eta$, i.e., a kinetic function raises the bound on $\HI$ by $K_*/\sqrt{\mathcal C_\nu}$, and a curvature coupling lowers the isocurvature by $d^2\mathcal C_\nu e^{-2(1-d)N_*}$.

\section{Kinetic functions}
\label{sec:slopes}
The results above assume a constant slope $p$, which is independent of the specific form of the kinetic function. For the linear and quadratic exponential functions we are interested in, $p$ is related to the inflaton trajectory by Eq.~\eqref{eq:plambdagamma}. For the linear exponential at nearly constant $\epsilon$, the slope $p_\lambda$ is nearly constant, and if the inflaton also generates the curvature perturbation, we find
\begin{equation}
\epsilon_*\simeq\frac{\HI^2}{8\pi^2\Mpl^2A_s},\qquad
\lambda\simeq-2p\,\frac{2\pi\sqrt{A_s}\,\Mpl}{\HI}.
\label{eq:lambdamap}
\end{equation}
A slope $p=-0.15$ then requires $\lambda\simeq2.1\times10^{4}$ at $\HI=10^{10}\GeV$ and $\lambda\simeq21$ at $10^{13}\GeV$. The large value at low scales reflects the small inflaton excursion, $\sqrt{2\epsilon}\,\Mpl$ per $e$-fold, over which $\ln K$ must change by $|p|$. Note that the relevant expansion parameter is $\lambda\phi/\Mpl$, which reaches $15.5$ at the pivot, and $K_\lambda$ must be read as a resummed kinetic function, such as the one provided by the hyperbolic field space of App.~\ref{app:sugra}, and not as its leading derivative operator. Eq.~(\ref{eq:lambdamap}) fixes $\lambda$ but not the potential, the scalar tilt, or the exit~\cite{Verner:2026cbn}. The high-scale case is the most natural one, and at $H_I = 10^{13} \, \rm{GeV}$ the tensor-to-scalar ratio is $r = 2 H_I^2/(\pi^2 M_{\rm Pl}^2 A_s) \simeq 1.6 \times 10^{-3}$, within the reach of next-generation $B$-mode searches~\cite{LiteBIRD:2022cnt}. A detection of primordial gravitational waves at this level would then point to the part of parameter space where a constant kinetic slope is easiest to realize, with a keV-scale ALP and a strongly blue isocurvature spectrum.

For the quadratic exponential we adopt an idealized constant $\epsilon$ trajectory that ends at the stationary point of $K_{\gamma}$, $\phi_e=0$, with $\phi\simeq\sqrt{2\epsilon}\,\Mpl N$ with $N=\ln(a_e/a)$, which gives
\begin{equation}
K_\gamma^2=e^{N^2/N_\gamma^2},\quad
p_\gamma=-\frac{N}{N_\gamma^2},\qquad
N_\gamma\equiv(2\epsilon\gamma)^{-1/2}.
\label{eq:gausshistory}
\end{equation}
Since the magnitude of the slope grows with the number of $e$-folds before the end of inflation, the spectrum is bluest on CMB scales, with local tilt $n_\eta(N)-1\simeq2N/N_\gamma^2$, and flatter on the scales that dominate the abundance. Since $|\dd p/\dd N|=N_\gamma^{-2}\ll1$, each mode leaves the horizon with the constant $p$ amplitude at its local slope. The kinetic coupling has $d=1$ throughout, and since $p=0$ at the exit the $q^{-2}$ tail of Eq.~\eqref{eq:UVoccupation} is absent.
Eq.~\eqref{eq:massiveIR} then applies mode by mode with
\begin{equation}
\mathcal P_{\eta,e}(q)\simeq\mathcal C_{\nu(N)}\left(\frac{\HI}{2\pi}\right)^2e^{-N^2/N_\gamma^2},
\qquad
\nu(N)=\frac32-\frac{N}{N_\gamma^2},
\label{eq:gaussspectrum}
\end{equation}
where $N=-\ln q$. Approximating $\mathcal C_\nu\simeq1$ and $\Ftr\simeq1$ below the turnover, and assuming $q_{\rm IR}<\sqrt\mu<q_{\rm max}$, we obtain the long mode estimate
\begin{equation}
Q_{\gamma,\rm long}\simeq\left(\frac{\HI}{2\pi}\right)^2\frac{\sqrt\pi\,N_\gamma}{2}\left[\operatorname{erf}\frac{N_{\rm IR}}{N_\gamma}-\operatorname{erf}\frac{N_{\rm osc}}{N_\gamma}\right],
\label{eq:Qgauss}
\end{equation}
with $N_{\rm IR}=-\ln q_{\rm IR}$ and $N_{\rm osc}=\frac12\ln(\HI/m_\eta)$, the number of $e$-folds before the end of inflation at which $k_{\rm osc}$ leaves the horizon.\footnote{Eq.~\eqref{eq:gaussspectrum} is the leading approximation to the Hankel-interval construction of App.~\ref{app:histories}. Eq.~\eqref{eq:Qgauss} additionally approximates the long mode transfer. For $N_\gamma\simeq15.6$, direct integration of the mode equation agrees with Eq.~\eqref{eq:gaussspectrum} to within $0.3\%$ over the tested range $11\leq N\leq52$, and at the $\HI=10^{10}\GeV$ threshold the running spectrum in Eq.~\eqref{eq:PX2} lowers the quadratic isocurvature power by $2.7\%$.}

Using the band, transfer, and isocurvature limit of Secs.~\ref{sec:abundance} and~\ref{sec:iso}, we find that the limit at $\HI=10^{10}\GeV$ requires $N_\gamma<15.6$, or a pivot tilt above $0.43$, with a threshold mass $m_\eta\simeq1.4\GeV$. This is more than two orders of magnitude below the $200\GeV$ found for a constant slope. The gain mostly comes from the running and the modes that set the abundance see a local tilt of only about $0.1$. The threshold corresponds to $\gamma\simeq2\times10^{7}$, and at $\HI=10^{11}$, $10^{12}$, and $10^{13}\GeV$ it moves to $N_\gamma<15.4$, $15.1$, and $14.7$, with masses $0.74\,$MeV, $0.59\,$keV, and $0.73\,$eV, respectively. At the tensor bound, $\HI=4.7\times10^{13}\GeV$, we find $N_\gamma<14.4$ and $m_\eta\simeq0.01\,$eV, and the required coupling falls from $\gamma\simeq23$ at $10^{13}\GeV$ to $\gamma\simeq1$.

Both kinetic functions can be realized without breaking the axion shift symmetry (App.~\ref{app:UV}). In supergravity with $Z$ the inflaton and $A$ the axion multiplet, a K\"ahler term $\mathcal Z_A(Z,\bar Z)(A+\bar A)^2$ reproduces either one along an $\alpha$-attractor trajectory, and the linear exponential also follows from a hyperbolic field space, $\mathcal K=-n\Mpl^2\ln(T+\bar T)$, with $|\lambda|=\sqrt{8/n}$. This field space must be strongly curved at low scales, and for $p=-0.15$ at $\HI=10^{10}\GeV$ we need $n\simeq1.8\times10^{-8}$, and a $T$-model at the same scale already requires an attractor parameter $\alpha\simeq4\times10^{-7}$. The renormalizable radial completion of App.~\ref{app:UV} gives $K^2=1+\gamma\phi^2/\Mpl^2$, and along the constant-$\epsilon$ trajectory its slope satisfies $|p|<1/N$, which is too small for the spectra of Sec.~\ref{sec:iso}. These constructions realize the kinetic functions but not the trajectory, the stabilization, or the exit, which must be specified separately, and none of them generates the curvature coupling $\xi$.

\section{Conclusions}
\label{sec:conclusions}
The answer to the question in our title is \emph{yes}, and the CMB sets the price. The two contributions can be compared directly because they are the same physics. We showed that modes still outside the horizon when the field starts to oscillate act as a local misalignment field, and the fluctuations carry the misalignment abundance of a displacement $\sqrt Q$, dominating when $Q>f_\eta^2\theta_c^2$. By matching the de~Sitter Hankel modes onto the radiation era parabolic cylinder modes, we obtained the late-time spectrum fully analytically, including the jump in $K'$ at the end of inflation, and the same calculation fixes the misalignment normalization without the order one ambiguity of an onset prescription.
 
The main constraint on this scenario comes from isocurvature, since the energy density is quadratic in the field and setting the mean to zero removes only the linear isocurvature term. The remaining quadratic term exceeds the Planck limit at the pivot by $8-9$ orders of magnitude for a flat or red spectrum, and a canonical axion therefore cannot be fluctuation-dominated dark matter. A blue tilt moves the power away from the scales probed by the CMB. However, a blue spectrum gives less abundance at a given mass and requires a heavier axion. For a constant kinetic slope the pivot limit requires $n_\eta-1\gtrsim0.27$ and a long-lived ALP heavier than about $200\GeV$ at $\HI=10^{10}\GeV$, and $n_\eta-1\gtrsim0.34$ with $m_\eta\gtrsim4.5\,$keV at $\HI=10^{13}\GeV$ (Fig.~\ref{fig:mstar}). The mean must also be small, and at $\HI=10^{10}\GeV$ it may supply at most $3\times10^{-5}$ of the abundance for $n_\eta-1=0.3$ and up to a third for $n_\eta-1=0.5$. A positive curvature coupling erases the mean dynamically, whereas for a kinetic function a mean as small as $|\theta_c|\lesssim4\times10^{-6}$ (for $f_\eta=10^{12}\GeV$ at $\HI=10^{10}\GeV$) must be assumed.

With a running slope these requirements become weaker, since the CMB constrains the tilt at the pivot, while the abundance is set tens of $e$-folds later, near $k_{\rm osc}$. The quadratic exponential is very blue at the pivot and only mildly blue near $k_{\rm osc}$. It lowers the required mass from $200\GeV$ to $1.4\GeV$ at $\HI=10^{10}\GeV$ and from $4.5\,$keV to $0.7\,$eV at $10^{13}\GeV$, and at the tensor bound it reaches $0.01\,$eV with an $\mathcal{O}(1)$ coupling, $\gamma\simeq1$.

The situation is less favorable for the QCD axion, as its mass is fixed by $f_\eta$ and it starts to oscillate late, near $T\sim1\GeV$. Within our estimate no single power law gives both the observed abundance and a small enough pivot isocurvature. A running tilt brings $\sqrt Q$ to about $2\times10^{11}\GeV$ near the tensor bound, close to the smallest misalignment displacement. However, if the PQ symmetry is restored near $T\simeq f_\eta$, keeping it broken after instantaneous reheating requires $f_\eta\gtrsim6\times10^{15}\GeV$, which raises $\eta_{\rm DM}$ to order $10^{13}\GeV$ and closes this window as well. A different PQ thermal history or a lower reheating temperature would require a separate analysis.
 
Realizing these kinetic functions at low inflationary scales requires large couplings or strongly curved field spaces, as the inflaton moves only $\sqrt{2\epsilon}\,\Mpl$ per $e$-fold. The linear exponential needs $\lambda\simeq2\times10^{4}$ at $\HI=10^{10}\GeV$ (falling to about $20$ at $10^{13}\GeV$), and its hyperbolic realization needs a strongly curved field space with $n\simeq2\times10^{-8}$. The high-scale case is the most natural one, and it predicts $r\simeq1.6\times10^{-3}$ at $\HI=10^{13}\GeV$, within the reach of next-generation $B$-mode searches. Both functions can be embedded in supergravity, but a renormalizable radial completion gives too small a slope, and none of these constructions fixes the inflaton trajectory or the exit. We leave two questions for future work. First, the blue non-Gaussian isocurvature should be confronted with a full CMB likelihood and not only with a limit at the pivot. Second, its growth on small scales, which is still allowed by Lyman-$\alpha$ and spectral distortion data, should be followed into early structure formation and induced gravitational waves~\cite{Domenech:2021ztg}, given that for $n_\eta-1=0.3$ at $\HI=10^{10}\GeV$ it exceeds the adiabatic power above $k\sim10^{3}\Mpc$.

\begin{acknowledgments}
We thank Ameen Ismail for his contributions in the early stages of this project, and Marcos A. G. García and Santiago Agüí Salcedo for useful discussions. S.J.L. is partly supported by the Samsung Science and Technology Foundation under
Project Number SSTF-BA2201-06.
The work of S.V. is supported by the Kavli Institute for Cosmological Physics at the University of Chicago. The work of B.Y. is partly supported by the NSF grant PHY-2309456.
\end{acknowledgments}

\appendix

\section{Derivation of the infrared spectrum}
\label{app:IR}
At $\tau_e=-1/\HI$ the small argument form $H^{(1)}_\nu(q)\simeq-(i/\pi)\Gamma(\nu)(2/q)^\nu$ leads to
\begin{equation}
|V_k|^2\simeq\frac{\mathcal C_\nu}{2\HI}\,q^{-2\nu},\qquad W_k\simeq d\,\HI V_k ,
\label{eq:Vsmall}
\end{equation}
where the second relation follows from $v_k^{\rm in}\propto(-\tau)^{1/2-\nu}$ and Eq.~\eqref{eq:jump}, with corrections of relative order $q^2$ for $\nu>1$. For $0<\nu<1$ corrections of order $q^{2\nu}$ also appear. The identity
\begin{equation}
D_a(0)=\frac{2^{a/2}\sqrt\pi}{\Gamma[(1-a)/2]}
\label{eq:D0}
\end{equation}
applied to Eq.~\eqref{eq:outmode} gives
\begin{equation}
|v_k^{\rm out}(0)|^2=\frac{\pi}{2\sqrt{m_\eta\HI}}\,\frac{e^{-\pi x^2/4}}{\left|\Gamma\!\left(\frac34+\frac{ix^2}{4}\right)\right|^2}.
\label{eq:vout0}
\end{equation}
For $q\ll1$ and $\mu\ll1$ the out mode at $y_e=1/\HI$ can be replaced by its value at $y=0$, with relative error of order $\max(q,\sqrt\mu)$, and the term $v_k^{{\rm out}\prime}V_k$ in Eq.~\eqref{eq:Bogoliubov} is smaller than $v_k^{\rm out}W_k$ by a factor of order $\max(q,\sqrt\mu)/d$. Keeping $v_k^{\rm out}W_k$ and using Eqs.~\eqref{eq:Vsmall} and~\eqref{eq:vout0} gives Eq.~\eqref{eq:massiveIR}. This approximation fails as $d\to0$, where both terms must be kept. At $p=0$, $\xi=1/6$, and exactly zero mass the in and out modes are plane waves and $\beta_k=0$. Note that the massive formula should not be evaluated by setting $m_\eta=0$ in its prefactor.

We evaluated Eq.~\eqref{eq:Bogoliubov} for $\mu=10^{-8}$ and $10^{-12}$, $0.01\leq x\leq30$ with $q\leq10^{-2}$, $p\in\{0,0.05,-0.15,-0.25\}$ at $\xi=0$, and $\xi\in\{0.020,0.036\}$ at $p=0$. The in and out modes satisfy their Wronskian conditions, $|\alpha_k|^2-|\beta_k|^2=1$ holds to working precision, and Eq.~\eqref{eq:massiveIR} agrees with the exact result to better than $2\times10^{-5}$ in this range. At conformal coupling the exact result follows $n_q\propto q^{-1}$ below the turnover. 

\section{Density correlator}
\label{app:density}
The produced modes are highly occupied and squeezed, so we treat the field as a classical Gaussian field in which each mode follows a real growing-mode transfer function, $\eta_{\bm k}(t)=G_k(t)\,\eta_{\bm k,e}$. On a uniform radiation slice, where $K=1$ and $R=0$, the density is quadratic in the field, and two modes with comoving momenta $\bm\ell$ and $\bm\ell'=\bm k-\bm\ell$ contribute through the kernel
\begin{equation}
\mathcal E_{\ell,\ell'}(t)=\frac12\,\overline{\dot G_\ell\dot G_{\ell'}+\left(m_\eta^2-\frac{\bm\ell\cdot\bm\ell'}{a^2}\right)G_\ell G_{\ell'}},
\label{eq:densitykernel}
\end{equation}
where the bar averages over the mass oscillations but not over the slower spatial evolution. The mean density and the connected density spectrum of the fluctuations are then
\begin{align}
\bar\rho_{\rm fluc}&=\int\frac{\dd^3\ell}{(2\pi)^3}\,\mathcal E_{\ell,-\ell}\,P_{\eta,e}(\ell),
\nonumber\\
P_{\rho,\rm fluc}(k)&=2\int\frac{\dd^3\ell}{(2\pi)^3}\,\mathcal E_{\ell,\ell'}^2\,P_{\eta,e}(\ell)\,P_{\eta,e}(\ell'),
\label{eq:fullconvolution}
\end{align}
with $P_{\eta,e}=2\pi^2\mathcal P_{\eta,e}/k^3$. The mean field adds a cross term with the fluctuations, which we include before dividing by the square of the total mean density.

Two limits matter. For $\ell,\ell'\ll k_{\rm osc}$ both modes are still outside the horizon when the oscillations begin and follow the same homogeneous solution, so the kernel reduces to a constant and Eq.~\eqref{eq:Wick} follows. At $k\ll k_{\rm osc}$, pairs with $\ell,\ell'\gtrsim k_{\rm osc}$ give a $k$-independent contribution to $P_{\rho,\rm fluc}$, which is white noise, while the long-mode contribution scales as $k^{3-4\nu}$. The ratio of the two is of order $(k/k_{\rm osc})^{4\nu-3}$. This classical treatment omits vacuum fluctuations, which a full calculation includes through the renormalized composite operator correlator~\cite{Herring:2019hbe}.

For the power law we measure comoving momenta in units of $a_e\HI$, so that $k=q$, and write $P_X(\ell)=2\pi^2\mathcal P_X(\ell)/\ell^3\propto\ell^{-2\nu}$. The convolution~\eqref{eq:PX2} then follows from
\begin{align}
\int\frac{\dd^3\ell}{(2\pi)^3}\,&\ell^{-2\nu}|\bm k-\bm\ell|^{-2\nu}
\nonumber\\
&=\frac{k^{3-4\nu}}{(4\pi)^{3/2}}\,
\frac{\Gamma^2\!\left(\frac32-\nu\right)\Gamma\!\left(2\nu-\frac32\right)}{\Gamma^2(\nu)\,\Gamma(3-2\nu)},
\label{eq:convolution}
\end{align}
valid for $3/4<\nu<3/2$, which gives $\mathcal P_{X^2}(q)=\mathcal D(\nu)\,\mathcal P_X^2(q)$ with $\mathcal D$ from Eq.~\eqref{eq:Dnu}. Restricting both momenta to $|\bm\ell|,|\bm k-\bm\ell|>q_{\rm IR}$ leaves an elementary angular integral. With $u=\ell/k$, $\varepsilon=q_{\rm IR}/k$, and $w=\max(|1-u|,\varepsilon)$,
\begin{equation}
\mathcal D_c(\nu;\varepsilon)=\frac{1}{2\nu-2}\int_\varepsilon^\infty\dd u\,u^{1-2\nu}\left[w^{2-2\nu}-(1+u)^{2-2\nu}\right].
\label{eq:Dc}
\end{equation}
This reduces to $\mathcal D(\nu)$ as $\varepsilon\to0$ and stays finite for $\nu\geq3/2$. At the pivot $\varepsilon=k_{\rm IR}/k_*=4.5\times10^{-3}$, independent of $\HI$. Because $\mathcal D_c$ depends on $k$ through $\varepsilon$, the tilt of $\PS$ exceeds $2(n_\eta-1)$ near the cutoff.

\section{Slowly varying kinetic histories}
\label{app:histories}
For a slowly varying $K(\tau)$ with $\xi=0$ we approximate $\ln K$ by linear pieces in $\ln a$, keeping $K$ continuous. In the de~Sitter interval $j$, with slope $p_j$, the basis modes are Hankel functions with $\nu_j=3/2+p_j$. At each boundary we match the field $\eta=v/z$ and its conjugate momentum $\pi_\eta=a^2K^2\eta'=zv'-z'v$, and we put them in a matrix form
\begin{equation}
\bm M_j(\tau)=
\begin{pmatrix}
v_j/z & v_j^*/z\\
zv_j'-z'v_j & zv_j^{*\prime}-z'v_j^*
\end{pmatrix},
\label{eq:Mmatrix}
\end{equation}
whose columns are the field and momentum of the normalized Hankel mode $v_j$ of interval $j$ and of its conjugate. The coefficients $(\alpha_j,\beta_j)$ of the mode in the basis $(v_j,v_j^*)$ then satisfy
\begin{equation}
\binom{\alpha_{j+1}}{\beta_{j+1}}=\bm M_{j+1}^{-1}(\tau_j)\,\bm M_j(\tau_j)\binom{\alpha_j}{\beta_j},
\label{eq:matrix}
\end{equation}
where $\tau_j$ is the boundary between intervals $j$ and $j+1$. Starting from $(\alpha_1,\beta_1)=(1,0)$ and ending in the radiation era basis~\eqref{eq:outmode}, we obtain $(\alpha_k,\beta_k)$ as a product of explicit matrices.

This extends the constant $p$ result to histories with a varying tilt, such as the quadratic exponential of Sec.~\ref{sec:slopes}, whose spectrum is blue across the CMB window and flatter near $k_{\rm osc}$. In the leading approximation each mode sees a single interval, with the slope at its horizon exit, which gives Eq.~\eqref{eq:gaussspectrum}. For a smooth history the result converges as the intervals are refined. Each boundary is a kink in $K$, however, and like the jump in $K'$ at the end of inflation (Sec.~\ref{sec:UVtail}) it excites spurious high-frequency modes, so a finite discretization is reliable only below a momentum cutoff that rises as the intervals are refined.

\section{Realizations of the kinetic function}
\label{app:UV}
This appendix gives explicit realizations of the kinetic functions $K_\lambda$ and $K_\gamma$ of Eq.~\eqref{eq:Kfunctions}.

\subsection{Radial realization}
\label{app:radial}
If $K$ comes from the radial mode, $K=\sigma_0(\phi)/f_\eta$. A renormalizable parent potential,
\begin{equation}
V_{\rm parent}=V_0(\phi)+\frac{\lambda_\sigma}{4}\left(\sigma^2-f_\eta^2-c_\sigma\phi^2\right)^2 ,
\label{eq:parent}
\end{equation}
gives on the broken branch
\begin{align}
\sigma_0^2(\phi)&=f_\eta^2+c_\sigma\phi^2, & K^2(\phi)&=1+\gamma\frac{\phi^2}{\Mpl^2},
\nonumber\\
\gamma&=c_\sigma\frac{\Mpl^2}{f_\eta^2}, & m_\sigma^2&=2\lambda_\sigma\sigma_0^2 .
\label{eq:radialUV}
\end{align}
A large $\gamma$ needs no large coupling and for $f_\eta=10^{11}\GeV$, $\gamma=2\times10^{7}$ corresponds to $c_\sigma\simeq3\times10^{-8}$. This is, however, only the leading term of the quadratic exponential. Along the trajectory of Sec.~\ref{sec:slopes} it gives $K^2=1+N^2/N_\gamma^2$ and $|p|=N^{-1}(N/N_\gamma)^2/[1+(N/N_\gamma)^2]<1/N$, too small to sustain $p=-0.15$ near $N\sim50$ and the higher terms of the quadratic exponential are what keep the slope growing. The blue spectra of Sec.~\ref{sec:iso} therefore need a nonpolynomial $\sigma_0(\phi)$ or a curved field space of the kind discussed next.

Integrating out $\sigma$ also corrects the inflaton kinetic term,
\begin{equation}
\mathcal L_{{\rm kin},\phi}=-\frac12\left[1+\left(\frac{\dd\sigma_0}{\dd\phi}\right)^2\right](\partial\phi)^2,
\label{eq:radialcorrection}
\end{equation}
which must remain small, $|\dd\sigma_0/\dd\phi|\lesssim1$. Since $\sigma_0$ changes by $|p|\,\sigma_0$ per $e$-fold while the canonical inflaton moves by $\sqrt{2\epsilon}\,\Mpl$, this requires
\begin{equation}
|p|\,Kf_\eta\lesssim\sqrt{2\epsilon}\,\Mpl=\frac{\HI}{2\pi\sqrt{A_s}},
\label{eq:radialbound}
\end{equation}
or $f_\eta\lesssim10^{11}\GeV$ at the pivot for $p=-0.15$ and $\HI=10^{10}\GeV$. Since $T_{\rm RH}\simeq8\times10^{13}\GeV$ there, the PQ sector must then avoid thermal restoration through weak couplings to the bath (Sec.~\ref{sec:setup}). The reduction also requires $m_\sigma\gg H,|\dot\sigma_0/\sigma_0|$ and physical momenta below $m_\sigma$, and loop corrections to $V_0$ must remain under control. A trajectory that approaches $\sigma_0=0$ invalidates the angular description before it produces an arbitrarily large enhancement.

\subsection{Supergravity}
\label{app:sugra}
Let $Z$ be a dimensionless inflaton multiplet, $S$ a stabilizer, and $A=(s+i\eta)/\sqrt2$ the axion multiplet, whose real part $s$ is the saxion. We take
\begin{align}
\mathcal K={}&-\frac{3\alpha}{2}\Mpl^2\ln\frac{(1-Z\bar Z)^2}{(1-Z^2)(1-\bar Z^2)}+S\bar S
\nonumber\\
&+\frac12\mathcal Z_A(Z,\bar Z)(A+\bar A)^2+\Delta\mathcal K_{\rm stab},
\label{eq:sugraK}\\
W={}&S\,\mathcal W(Z)+W_{\rm br}(A).
\label{eq:sugraW}
\end{align}
The first term of $\mathcal K$ describes the hyperbolic disk, $\mathcal K_{Z\bar Z}=3\alpha\Mpl^2/(1-Z\bar Z)^2$, as in Ref.~\cite{Carrasco:2015rva}. The $F$-term potential is given by
\begin{equation}
V_F=e^{\mathcal K/\Mpl^2}\left(\mathcal K^{I\bar J}D_IW\,D_{\bar J}\bar W-\frac{3|W|^2}{\Mpl^2}\right),
\label{eq:VF}
\end{equation}
with $D_IW=\partial_IW+\mathcal K_IW/\Mpl^2$. Along $S=s=0$ and $Z=\bar Z$, and neglecting the small term $W_{\rm br}$, we have $\mathcal K=W=0$ and $\mathcal K_{S\bar S}=1$, so that
\begin{equation}
Z=\tanh\frac{\phi}{\sqrt{6\alpha}\,\Mpl},\qquad V=|\mathcal W(Z)|^2 .
\label{eq:sugratraj}
\end{equation}
A linear $\mathcal W\propto Z$ gives the $T$-model~\cite{Kallosh:2013yoa,Carrasco:2015rva, Ellis:2019bmm, Ellis:2021kad, Ellis:2025zrf}. The real function
\begin{equation}
\hat\phi(Z,\bar Z)=\sqrt{\frac{3\alpha}{8}}\,\Mpl\ln\frac{(1+Z)(1+\bar Z)}{(1-Z)(1-\bar Z)}
\label{eq:phihat}
\end{equation}
equals $\phi$ on the trajectory, and the choices
\begin{equation}
\mathcal Z_A=e^{\lambda\hat\phi/\Mpl}
\quad\text{or}\quad
\mathcal Z_A=e^{\gamma\hat\phi^2/\Mpl^2}
\label{eq:ZAsugra}
\end{equation}
give $\mathcal K_{A\bar A}=K^2(\phi)$ there. Because $\mathcal Z_A$ multiplies $(A+\bar A)^2$, it preserves the axion shift symmetry until $W_{\rm br}$ is included, and it leaves the inflationary trajectory unchanged.

Terms such as~\cite{Kallosh:2010xz}
\begin{align}
\Delta\mathcal K_{\rm stab}={}&-c_S\frac{(S\bar S)^2}{\Mpl^2}-c_A\frac{S\bar S\,\mathcal Z_A(A+\bar A)^2}{\Mpl^2}
\nonumber\\
&-c_Z\frac{S\bar S\,\mathcal Y^2}{\Mpl^2},
\label{eq:stabilization}
\end{align}
with $\mathcal Y=\sqrt2\,{\rm Im}[\sqrt{6\alpha}\,\Mpl\operatorname{artanh}Z]$, can stabilize the transverse directions. They vanish on the trajectory and, for positive coefficients, raise the masses of the transverse fields. Their coefficients must be fixed by the full mass matrix and the cutoff, and small $\alpha$ is not automatically stable. The axion potential, the exit from inflation, and the saxion dynamics remain separate model building questions.

The linear exponential also follows directly from the geometry, without a choice of $\mathcal Z_A$. For $\mathcal K=-n\Mpl^2\ln(T+\bar T)$ and $u\equiv{\rm Re}\,T$, the fields $\phi=\sqrt{n/2}\,\Mpl\ln(u/u_0)$ and $\eta=\sqrt{n/2}\,\Mpl\,{\rm Im}\,T/u_0$, with constant $u_0$, have kinetic terms
\begin{equation}
\mathcal L_{\rm kin}=-\frac12(\partial\phi)^2-\frac12e^{-\sqrt{8/n}\,\phi/\Mpl}(\partial\eta)^2 ,
\label{eq:hyperbolic}
\end{equation}
so $|\lambda|=\sqrt{8/n}$, with the sign set by the direction of the roll, and $|p|=2\sqrt{\epsilon/n}$. The example $p=-0.15$ at $\HI=10^{10}\GeV$ requires $n\simeq1.8\times10^{-8}$, a small field-space curvature radius. For comparison, a $T$-model at this scale needs $\alpha\simeq4\times10^{-7}$. This geometry is related to the one used in Ref.~\cite{Tadepalli:2026mdc} to obtain blue axion spectra.

These constructions realize $K(\phi)$ locally but do not generate the shift-breaking curvature term in Eq.~\eqref{eq:action}. For $\xi\neq0$ that term, and any terms induced by a Jordan to Einstein frame transformation, must be specified separately.

\bibliographystyle{apsrev4-2}
\bibliography{references}

@article{Preskill:1982cy,
    author = "Preskill, John and Wise, Mark B. and Wilczek, Frank",
    editor = "Srednicki, M. A.",
    title = "{Cosmology of the Invisible Axion}",
    reportNumber = "HUTP-82-A048, NSF-ITP-82-103",
    doi = "10.1016/0370-2693(83)90637-8",
    journal = "Phys. Lett. B",
    volume = "120",
    pages = "127--132",
    year = "1983"
}

@article{Abbott:1982af,
    author = "Abbott, L. F. and Sikivie, P.",
    editor = "Srednicki, M. A.",
    title = "{A Cosmological Bound on the Invisible Axion}",
    reportNumber = "PRINT-82-0695 (BRANDEIS)",
    doi = "10.1016/0370-2693(83)90638-X",
    journal = "Phys. Lett. B",
    volume = "120",
    pages = "133--136",
    year = "1983"
}

@article{Dine:1982ah,
    author = "Dine, Michael and Fischler, Willy",
    editor = "Srednicki, M. A.",
    title = "{The Not So Harmless Axion}",
    reportNumber = "UPR-0201T",
    doi = "10.1016/0370-2693(83)90639-1",
    journal = "Phys. Lett. B",
    volume = "120",
    pages = "137--141",
    year = "1983"
}

@article{Peccei:1977hh,
    author = "Peccei, R. D. and Quinn, Helen R.",
    title = "{CP Conservation in the Presence of Instantons}",
    reportNumber = "ITP-568-STANFORD",
    doi = "10.1103/PhysRevLett.38.1440",
    journal = "Phys. Rev. Lett.",
    volume = "38",
    pages = "1440--1443",
    year = "1977"
}

@article{Peccei:1977ur,
    author = "Peccei, R. D. and Quinn, Helen R.",
    title = "{Constraints Imposed by CP Conservation in the Presence of Instantons}",
    reportNumber = "ITP-572-STANFORD",
    doi = "10.1103/PhysRevD.16.1791",
    journal = "Phys. Rev. D",
    volume = "16",
    pages = "1791--1797",
    year = "1977"
}

@article{Weinberg:1977ma,
    author = "Weinberg, Steven",
    title = "{A New Light Boson?}",
    reportNumber = "HUTP-77/A074",
    doi = "10.1103/PhysRevLett.40.223",
    journal = "Phys. Rev. Lett.",
    volume = "40",
    pages = "223--226",
    year = "1978"
}

@article{Wilczek:1977pj,
    author = "Wilczek, Frank",
    title = "{Problem of Strong  $P$  and  $T$  Invariance in the Presence of Instantons}",
    reportNumber = "Print-77-0939 (COLUMBIA)",
    doi = "10.1103/PhysRevLett.40.279",
    journal = "Phys. Rev. Lett.",
    volume = "40",
    pages = "279--282",
    year = "1978"
}

@article{Axenides:1983hj,
    author = "Axenides, Minos and Brandenberger, Robert H. and Turner, Michael S.",
    title = "{Development of Axion Perturbations in an Axion Dominated Universe}",
    reportNumber = "HUTP-83/A021, HUTMP B132",
    doi = "10.1016/0370-2693(83)90586-5",
    journal = "Phys. Lett. B",
    volume = "126",
    pages = "178--182",
    year = "1983"
}

@article{Seckel:1985tj,
    author = "Seckel, D. and Turner, Michael S.",
    title = "{Isothermal Density Perturbations in an Axion Dominated Inflationary Universe}",
    reportNumber = "FERMILAB-PUB-85-087-A",
    doi = "10.1103/PhysRevD.32.3178",
    journal = "Phys. Rev. D",
    volume = "32",
    pages = "3178",
    year = "1985"
}

@article{Linde:1985yf,
    author = "Linde, Andrei D.",
    title = "{Generation of Isothermal Density Perturbations in the Inflationary Universe}",
    reportNumber = "LEBEDEV-85-65",
    doi = "10.1016/0370-2693(85)90436-8",
    journal = "Phys. Lett. B",
    volume = "158",
    pages = "375--380",
    year = "1985"
}

@article{Lyth:1989pb,
    author = "Lyth, David H.",
    title = "{A Limit on the Inflationary Energy Density From Axion Isocurvature Fluctuations}",
    reportNumber = "LANCASTER-89TH10",
    doi = "10.1016/0370-2693(90)90374-F",
    journal = "Phys. Lett. B",
    volume = "236",
    pages = "408--410",
    year = "1990"
}

@article{Turner:1990uz,
    author = "Turner, Michael S. and Wilczek, Frank",
    title = "{Inflationary Axion Cosmology}",
    reportNumber = "FERMILAB-PUB-90-198-A",
    doi = "10.1103/PhysRevLett.66.5",
    journal = "Phys. Rev. Lett.",
    volume = "66",
    pages = "5--8",
    year = "1991"
}

@article{Linde:1991km,
    author = "Linde, Andrei D.",
    title = "{Axions in inflationary cosmology}",
    reportNumber = "SU-ITP-883",
    doi = "10.1016/0370-2693(91)90130-I",
    journal = "Phys. Lett. B",
    volume = "259",
    pages = "38--47",
    year = "1991"
}

@article{Lyth:1992tx,
    author = "Lyth, David H. and Stewart, Ewan D.",
    title = "{Constraining the inflationary energy scale from axion cosmology}",
    reportNumber = "LANC-TH-92-04",
    doi = "10.1016/0370-2693(92)90006-P",
    journal = "Phys. Lett. B",
    volume = "283",
    pages = "189--193",
    year = "1992"
}

@article{Kobayashi:2016qld,
    author = "Kobayashi, Takeshi and Takahashi, Fuminobu",
    title = "{Cosmological Perturbations of Axion with a Dynamical Decay Constant}",
    eprint = "1607.04294",
    archivePrefix = "arXiv",
    primaryClass = "hep-ph",
    reportNumber = "SISSA-39-2016-FISI, TU-1026, IPMU16-0118",
    doi = "10.1088/1475-7516/2016/08/056",
    journal = "JCAP",
    volume = "08",
    pages = "056",
    year = "2016"
}

@misc{Freese:2026xax,
    author = "Freese, Katherine and Sfakianakis, Evangelos I. and Shams Es Haghi, Barmak",
    title = "{Isocurvature-Free QCD Axion Dark Matter from Inflaton-Driven Early QCD: the Necessity of Inflationary Plateaus}",
    eprint = "2605.15192",
    archivePrefix = "arXiv",
    primaryClass = "astro-ph.CO",
    month = "5",
    year = "2026"
}

@misc{Fox:2004kb,
    author = "Fox, Patrick and Pierce, Aaron and Thomas, Scott D.",
    title = "{Probing a QCD string axion with precision cosmological measurements}",
    eprint = "hep-th/0409059",
    archivePrefix = "arXiv",
    reportNumber = "SLAC-PUB-10030, SU-ITP-03-19, SCIPP-2003-04",
    month = "9",
    year = "2004"
}

@article{Hertzberg:2008wr,
    author = "Hertzberg, Mark P and Tegmark, Max and Wilczek, Frank",
    title = "{Axion Cosmology and the Energy Scale of Inflation}",
    eprint = "0807.1726",
    archivePrefix = "arXiv",
    primaryClass = "astro-ph",
    reportNumber = "MIT-CTP-3950",
    doi = "10.1103/PhysRevD.78.083507",
    journal = "Phys. Rev. D",
    volume = "78",
    pages = "083507",
    year = "2008"
}

@article{Planck:2018jri,
    author = "Akrami, Y. and others",
    collaboration = "Planck",
    title = "{Planck 2018 results. X. Constraints on inflation}",
    eprint = "1807.06211",
    archivePrefix = "arXiv",
    primaryClass = "astro-ph.CO",
    doi = "10.1051/0004-6361/201833887",
    journal = "Astron. Astrophys.",
    volume = "641",
    pages = "A10",
    year = "2020"
}

@article{Kearney:2016vqw,
    author = "Kearney, John and Orlofsky, Nicholas and Pierce, Aaron",
    title = "{High-Scale Axions without Isocurvature from Inflationary Dynamics}",
    eprint = "1601.03049",
    archivePrefix = "arXiv",
    primaryClass = "hep-ph",
    reportNumber = "FERMILAB-PUB-16-005-T, MCTP-16-02",
    doi = "10.1103/PhysRevD.93.095026",
    journal = "Phys. Rev. D",
    volume = "93",
    number = "9",
    pages = "095026",
    year = "2016"
}

@article{Graham:2025iwx,
    author = "Graham, Peter W. and Racco, Davide",
    title = "{Revisiting isocurvature bounds on the minimal QCD axion}",
    eprint = "2506.03348",
    archivePrefix = "arXiv",
    primaryClass = "hep-ph",
    reportNumber = "ZU-TH 38/25",
    doi = "10.1007/JHEP12(2025)028",
    journal = "JHEP",
    volume = "12",
    pages = "028",
    year = "2025"
}

@article{Berbig:2024ufe,
    author = "Berbig, Maximilian",
    title = "{Minimal solution to the axion isocurvature problem from nonminimal coupling}",
    eprint = "2404.06441",
    archivePrefix = "arXiv",
    primaryClass = "hep-ph",
    doi = "10.1103/PhysRevD.110.095008",
    journal = "Phys. Rev. D",
    volume = "110",
    number = "9",
    pages = "095008",
    year = "2024"
}

@misc{Verner:2026cbn,
    author = "Verner, Sarunas",
    title = "{Axion isocurvature and the model-building problem of low-scale inflation}",
    eprint = "2607.22809",
    archivePrefix = "arXiv",
    primaryClass = "hep-ph",
    month = "7",
    year = "2026"
}

@article{Parker:1969au,
    author = "Parker, Leonard",
    title = "{Quantized fields and particle creation in expanding universes. 1.}",
    doi = "10.1103/PhysRev.183.1057",
    journal = "Phys. Rev.",
    volume = "183",
    pages = "1057--1068",
    year = "1969"
}

@article{Ford:1986sy,
    author = "Ford, L. H.",
    title = "{Gravitational Particle Creation and Inflation}",
    reportNumber = "TUTP-86-8",
    doi = "10.1103/PhysRevD.35.2955",
    journal = "Phys. Rev. D",
    volume = "35",
    pages = "2955",
    year = "1987"
}

@book{Birrell:1982ix,
    author = "Birrell, N. D. and Davies, P. C. W.",
    title = "{Quantum Fields in Curved Space}",
    doi = "10.1017/CBO9780511622632",
    isbn = "978-0-511-62263-2, 978-0-521-27858-4",
    publisher = "Cambridge University Press",
    address = "Cambridge, UK",
    series = "Cambridge Monographs on Mathematical Physics",
    year = "1982"
}

@misc{Dvali:1995ce,
    author = "Dvali, G. R.",
    title = "{Removing the cosmological bound on the axion scale}",
    eprint = "hep-ph/9505253",
    archivePrefix = "arXiv",
    reportNumber = "IFUP-TH-21-95",
    month = "5",
    year = "1995"
}

@article{Kitano:2023mra,
    author = "Kitano, Ryuichiro and Suzuki, Motoo and Yin, Wen",
    title = "{Role of QCD in moduli stabilization during inflation and axion dark matter}",
    eprint = "2307.15059",
    archivePrefix = "arXiv",
    primaryClass = "hep-ph",
    reportNumber = "TU-1195",
    doi = "10.1007/JHEP11(2023)103",
    journal = "JHEP",
    volume = "11",
    pages = "103",
    year = "2023"
}

@article{Kaneta:2022gug,
    author = "Kaneta, Kunio and Lee, Sung Mook and Oda, Kin-ya",
    title = "{Boltzmann or Bogoliubov? Approaches compared in gravitational particle production}",
    eprint = "2206.10929",
    archivePrefix = "arXiv",
    primaryClass = "astro-ph.CO",
    doi = "10.1088/1475-7516/2022/09/018",
    journal = "JCAP",
    volume = "09",
    pages = "018",
    year = "2022"
}

@article{Kolb:2023ydq,
    author = "Kolb, Edward W. and Long, Andrew J.",
    title = "{Cosmological gravitational particle production and its implications for cosmological relics}",
    eprint = "2312.09042",
    archivePrefix = "arXiv",
    primaryClass = "astro-ph.CO",
    doi = "10.1103/RevModPhys.96.045005",
    journal = "Rev. Mod. Phys.",
    volume = "96",
    number = "4",
    pages = "045005",
    year = "2024"
}

@article{Chung:1998zb,
    author = "Chung, Daniel J. H. and Kolb, Edward W. and Riotto, Antonio",
    title = "{Superheavy dark matter}",
    eprint = "hep-ph/9802238",
    archivePrefix = "arXiv",
    reportNumber = "FERMILAB-PUB-98-021-A, CERN-TH-98-37, OUTP-98-02-P, FERMILAB-FERMILAB-PUB-98-021-A, CERN-CERN-TH-98-37, OXFORD-U. --OUTP-98-02-P",
    doi = "10.1103/PhysRevD.59.023501",
    journal = "Phys. Rev. D",
    volume = "59",
    pages = "023501",
    year = "1998"
}

@article{Ema:2018ucl,
    author = "Ema, Yohei and Nakayama, Kazunori and Tang, Yong",
    title = "{Production of Purely Gravitational Dark Matter}",
    eprint = "1804.07471",
    archivePrefix = "arXiv",
    primaryClass = "hep-ph",
    reportNumber = "UT-18-08, KEK-TH-2047",
    doi = "10.1007/JHEP09(2018)135",
    journal = "JHEP",
    volume = "09",
    pages = "135",
    year = "2018"
}

@article{Ling:2021zlj,
    author = "Ling, Siyang and Long, Andrew J.",
    title = "{Superheavy scalar dark matter from gravitational particle production in $\alpha$-attractor models of inflation}",
    eprint = "2101.11621",
    archivePrefix = "arXiv",
    primaryClass = "astro-ph.CO",
    doi = "10.1103/PhysRevD.103.103532",
    journal = "Phys. Rev. D",
    volume = "103",
    number = "10",
    pages = "103532",
    year = "2021"
}

@article{Racco:2024aac,
    author = "Racco, Davide and Verner, Sarunas and Xue, Wei",
    title = "{Gravitational production of heavy particles during and after inflation}",
    eprint = "2405.13883",
    archivePrefix = "arXiv",
    primaryClass = "hep-ph",
    doi = "10.1007/JHEP09(2024)129",
    journal = "JHEP",
    volume = "09",
    pages = "129",
    year = "2024"
}

@article{Verner:2024agh,
    author = "Verner, Sarunas",
    title = "{Nonminimal superheavy dark matter}",
    eprint = "2408.11889",
    archivePrefix = "arXiv",
    primaryClass = "hep-ph",
    doi = "10.1088/1475-7516/2025/05/060",
    journal = "JCAP",
    volume = "05",
    pages = "060",
    year = "2025"
}

@article{Garcia:2023awt,
    author = "Garcia, Marcos A. G. and Pierre, Mathias and Verner, Sarunas",
    title = "{Isocurvature constraints on scalar dark matter production from the inflaton}",
    eprint = "2303.07359",
    archivePrefix = "arXiv",
    primaryClass = "hep-ph",
    reportNumber = "DESY-23-027",
    doi = "10.1103/PhysRevD.107.123508",
    journal = "Phys. Rev. D",
    volume = "107",
    number = "12",
    pages = "123508",
    year = "2023"
}

@article{Garcia:2023qab,
    author = "Garcia, Marcos A. G. and Pierre, Mathias and Verner, Sarunas",
    title = "{New window into gravitationally produced scalar dark matter}",
    eprint = "2305.14446",
    archivePrefix = "arXiv",
    primaryClass = "hep-ph",
    reportNumber = "DESY-23-065",
    doi = "10.1103/PhysRevD.108.115024",
    journal = "Phys. Rev. D",
    volume = "108",
    number = "11",
    pages = "115024",
    year = "2023"
}

@article{Herring:2019hbe,
    author = "Herring, Nathan and Boyanovsky, Daniel and Zentner, Andrew R.",
    title = "{Nonadiabatic cosmological production of ultralight dark matter}",
    eprint = "1912.10859",
    archivePrefix = "arXiv",
    primaryClass = "gr-qc",
    doi = "10.1103/PhysRevD.101.083516",
    journal = "Phys. Rev. D",
    volume = "101",
    number = "8",
    pages = "083516",
    year = "2020"
}

@article{Starobinsky:1994bd,
    author = "Starobinsky, Alexei A. and Yokoyama, Junichi",
    title = "{Equilibrium state of a selfinteracting scalar field in the De Sitter background}",
    eprint = "astro-ph/9407016",
    archivePrefix = "arXiv",
    reportNumber = "YITP-U-94-12",
    doi = "10.1103/PhysRevD.50.6357",
    journal = "Phys. Rev. D",
    volume = "50",
    pages = "6357--6368",
    year = "1994"
}

@article{Graham:2018jyp,
    author = "Graham, Peter W. and Scherlis, Adam",
    title = "{Stochastic axion scenario}",
    eprint = "1805.07362",
    archivePrefix = "arXiv",
    primaryClass = "hep-ph",
    doi = "10.1103/PhysRevD.98.035017",
    journal = "Phys. Rev. D",
    volume = "98",
    number = "3",
    pages = "035017",
    year = "2018"
}

@article{Takahashi:2018tdu,
    author = "Takahashi, Fuminobu and Yin, Wen and Guth, Alan H.",
    title = "{QCD axion window and low-scale inflation}",
    eprint = "1805.08763",
    archivePrefix = "arXiv",
    primaryClass = "hep-ph",
    reportNumber = "MIT-CTP/5022, TU-1064, IPMU18-0088, MIT-CTP-5022",
    doi = "10.1103/PhysRevD.98.015042",
    journal = "Phys. Rev. D",
    volume = "98",
    number = "1",
    pages = "015042",
    year = "2018"
}

@article{Markkanen:2018gcw,
    author = "Markkanen, Tommi and Rajantie, Arttu and Tenkanen, Tommi",
    title = "{Spectator Dark Matter}",
    eprint = "1811.02586",
    archivePrefix = "arXiv",
    primaryClass = "astro-ph.CO",
    reportNumber = "IMPERIAL/TP/2018/TM/06",
    doi = "10.1103/PhysRevD.98.123532",
    journal = "Phys. Rev. D",
    volume = "98",
    number = "12",
    pages = "123532",
    year = "2018"
}

@article{Tenkanen:2019aij,
    author = "Tenkanen, Tommi",
    title = "{Dark matter from scalar field fluctuations}",
    eprint = "1905.01214",
    archivePrefix = "arXiv",
    primaryClass = "astro-ph.CO",
    doi = "10.1103/PhysRevLett.123.061302",
    journal = "Phys. Rev. Lett.",
    volume = "123",
    number = "6",
    pages = "061302",
    year = "2019"
}

@article{Nurmi:2015ema,
    author = "Nurmi, Sami and Tenkanen, Tommi and Tuominen, Kimmo",
    title = "{Inflationary Imprints on Dark Matter}",
    eprint = "1506.04048",
    archivePrefix = "arXiv",
    primaryClass = "astro-ph.CO",
    reportNumber = "HIP-2015-6-TH",
    doi = "10.1088/1475-7516/2015/11/001",
    journal = "JCAP",
    volume = "11",
    pages = "001",
    year = "2015"
}

@article{Garcia:2025rut,
    author = "Garcia, Marcos A. G. and Ke, Wenqi and Mambrini, Yann and Olive, Keith A. and Verner, Sarunas",
    title = "{Scalar field fluctuations and the production of dark matter}",
    eprint = "2502.20471",
    archivePrefix = "arXiv",
    primaryClass = "hep-ph",
    reportNumber = "UMN--TH--4417/25, FTPI--MINN--25/02",
    doi = "10.1088/1475-7516/2025/08/039",
    journal = "JCAP",
    volume = "08",
    pages = "039",
    year = "2025"
}

@article{Graham:2015rva,
    author = "Graham, Peter W. and Mardon, Jeremy and Rajendran, Surjeet",
    title = "{Vector Dark Matter from Inflationary Fluctuations}",
    eprint = "1504.02102",
    archivePrefix = "arXiv",
    primaryClass = "hep-ph",
    doi = "10.1103/PhysRevD.93.103520",
    journal = "Phys. Rev. D",
    volume = "93",
    number = "10",
    pages = "103520",
    year = "2016"
}

@article{Ema:2019yrd,
    author = "Ema, Yohei and Nakayama, Kazunori and Tang, Yong",
    title = "{Production of purely gravitational dark matter: the case of fermion and vector boson}",
    eprint = "1903.10973",
    archivePrefix = "arXiv",
    primaryClass = "hep-ph",
    reportNumber = "DESY-19-050, DESY 19-050, KEK-TH-2114, UT-19-04",
    doi = "10.1007/JHEP07(2019)060",
    journal = "JHEP",
    volume = "07",
    pages = "060",
    year = "2019"
}

@article{Ahmed:2020fhc,
    author = "Ahmed, Aqeel and Grzadkowski, Bohdan and Socha, Anna",
    title = "{Gravitational production of vector dark matter}",
    eprint = "2005.01766",
    archivePrefix = "arXiv",
    primaryClass = "hep-ph",
    doi = "10.1007/JHEP08(2020)059",
    journal = "JHEP",
    volume = "08",
    pages = "059",
    year = "2020"
}

@article{Kolb:2020fwh,
    author = "Kolb, Edward W. and Long, Andrew J.",
    title = "{Completely dark photons from gravitational particle production during the inflationary era}",
    eprint = "2009.03828",
    archivePrefix = "arXiv",
    primaryClass = "astro-ph.CO",
    doi = "10.1007/JHEP03(2021)283",
    journal = "JHEP",
    volume = "03",
    pages = "283",
    year = "2021"
}

@article{Ratra:1991bn,
    author = "Ratra, Bharat",
    title = "{Cosmological 'seed' magnetic field from inflation}",
    reportNumber = "CALT-68-1750, GRP-286",
    doi = "10.1086/186384",
    journal = "Astrophys. J. Lett.",
    volume = "391",
    pages = "L1--L4",
    year = "1992"
}

@article{Martin:2007ue,
    author = "Martin, Jerome and Yokoyama, Jun'ichi",
    title = "{Generation of Large-Scale Magnetic Fields in Single-Field Inflation}",
    eprint = "0711.4307",
    archivePrefix = "arXiv",
    primaryClass = "astro-ph",
    doi = "10.1088/1475-7516/2008/01/025",
    journal = "JCAP",
    volume = "01",
    pages = "025",
    year = "2008"
}

@article{Nakai:2020cfw,
    author = "Nakai, Yuichiro and Namba, Ryo and Wang, Ziwei",
    title = "{Light Dark Photon Dark Matter from Inflation}",
    eprint = "2004.10743",
    archivePrefix = "arXiv",
    primaryClass = "hep-ph",
    doi = "10.1007/JHEP12(2020)170",
    journal = "JHEP",
    volume = "12",
    pages = "170",
    year = "2020"
}

@article{Nakayama:2020rka,
    author = "Nakayama, Kazunori",
    title = "{Constraint on Vector Coherent Oscillation Dark Matter with Kinetic Function}",
    eprint = "2004.10036",
    archivePrefix = "arXiv",
    primaryClass = "hep-ph",
    reportNumber = "UT-20-04",
    doi = "10.1088/1475-7516/2020/08/033",
    journal = "JCAP",
    volume = "08",
    pages = "033",
    year = "2020"
}

@misc{Ismail:2024zbq,
    author = "Ismail, Ameen and Lee, Seung J. and Yu, Bingrong",
    title = "{Axion dark matter from inflation-driven quantum phase transition}",
    eprint = "2402.08716",
    archivePrefix = "arXiv",
    primaryClass = "hep-ph",
    month = "2",
    year = "2024"
}

@article{Cremmer:1982en,
    author = "Cremmer, E. and Ferrara, S. and Girardello, L. and Van Proeyen, Antoine",
    editor = "Salam, A. and Sezgin, E.",
    title = "{Yang-Mills Theories with Local Supersymmetry: Lagrangian, Transformation Laws and SuperHiggs Effect}",
    reportNumber = "CERN-TH-3348",
    doi = "10.1016/0550-3213(83)90679-X",
    journal = "Nucl. Phys. B",
    volume = "212",
    pages = "413",
    year = "1983"
}

@article{Svrcek:2006yi,
    author = "Svrcek, Peter and Witten, Edward",
    title = "{Axions In String Theory}",
    eprint = "hep-th/0605206",
    archivePrefix = "arXiv",
    reportNumber = "SLAC-PUB-11894",
    doi = "10.1088/1126-6708/2006/06/051",
    journal = "JHEP",
    volume = "06",
    pages = "051",
    year = "2006"
}

@article{Tadepalli:2026mdc,
    author = "Tadepalli, Sai Chaitanya",
    title = "{Geometric suppression of CMB isocurvature and a blue-tilted spectrum}",
    eprint = "2601.22221",
    archivePrefix = "arXiv",
    primaryClass = "hep-ph",
    doi = "10.1103/fpd7-xdhl",
    journal = "Phys. Rev. D",
    volume = "114",
    number = "2",
    pages = "023550",
    year = "2026"
}

@article{Jeong:2013xta,
    author = "Jeong, Kwang Sik and Takahashi, Fuminobu",
    title = "{Suppressing Isocurvature Perturbations of QCD Axion Dark Matter}",
    eprint = "1304.8131",
    archivePrefix = "arXiv",
    primaryClass = "hep-ph",
    reportNumber = "TU-935",
    doi = "10.1016/j.physletb.2013.10.061",
    journal = "Phys. Lett. B",
    volume = "727",
    pages = "448--451",
    year = "2013"
}

@article{Higaki:2014ooa,
    author = "Higaki, Tetsutaro and Jeong, Kwang Sik and Takahashi, Fuminobu",
    title = "{Solving the Tension between High-Scale Inflation and Axion Isocurvature Perturbations}",
    eprint = "1403.4186",
    archivePrefix = "arXiv",
    primaryClass = "hep-ph",
    reportNumber = "KEK-TH-1711, DESY-14-033, TU-958, IPMU14-0060",
    doi = "10.1016/j.physletb.2014.05.014",
    journal = "Phys. Lett. B",
    volume = "734",
    pages = "21--26",
    year = "2014"
}

@article{Folkerts:2013tua,
    author = "Folkerts, Sarah and Germani, Cristiano and Redondo, Javier",
    title = "{Axion Dark Matter and Planck favor non-minimal couplings to gravity}",
    eprint = "1304.7270",
    archivePrefix = "arXiv",
    primaryClass = "hep-ph",
    reportNumber = "LMU-ASC-26-13",
    doi = "10.1016/j.physletb.2013.12.026",
    journal = "Phys. Lett. B",
    volume = "728",
    pages = "532--536",
    year = "2014"
}

@article{Kasuya:2009up,
    author = "Kasuya, Shinta and Kawasaki, Masahiro",
    title = "{Axion isocurvature fluctuations with extremely blue spectrum}",
    eprint = "0904.3800",
    archivePrefix = "arXiv",
    primaryClass = "astro-ph.CO",
    doi = "10.1103/PhysRevD.80.023516",
    journal = "Phys. Rev. D",
    volume = "80",
    pages = "023516",
    year = "2009"
}

@article{Chung:2015pga,
    author = "Chung, Daniel J. H. and Yoo, Hojin",
    title = "{Elementary Theorems Regarding Blue Isocurvature Perturbations}",
    eprint = "1501.05618",
    archivePrefix = "arXiv",
    primaryClass = "astro-ph.CO",
    doi = "10.1103/PhysRevD.91.083530",
    journal = "Phys. Rev. D",
    volume = "91",
    pages = "083530",
    year = "2015"
}

@article{Chung:2016wvv,
    author = "Chung, Daniel J. H. and Upadhye, Amol",
    title = "{Bump in the blue axion isocurvature spectrum}",
    eprint = "1610.04284",
    archivePrefix = "arXiv",
    primaryClass = "astro-ph.CO",
    doi = "10.1103/PhysRevD.95.023503",
    journal = "Phys. Rev. D",
    volume = "95",
    number = "2",
    pages = "023503",
    year = "2017"
}

@article{Kalia:2025uxg,
    author = "Kalia, Saarik",
    title = "{Dynamically generated tilt of isocurvature fluctuations}",
    eprint = "2510.11803",
    archivePrefix = "arXiv",
    primaryClass = "hep-ph",
    reportNumber = "UMN-TH-4510/25",
    doi = "10.1103/q86r-7msm",
    journal = "Phys. Rev. D",
    volume = "114",
    number = "2",
    pages = "023505",
    year = "2026"
}

@article{Alonso-Alvarez:2018tus,
    author = "Alonso-{\'A}lvarez, Gonzalo and Jaeckel, Joerg",
    title = "{Lightish but clumpy: scalar dark matter from inflationary fluctuations}",
    eprint = "1807.09785",
    archivePrefix = "arXiv",
    primaryClass = "hep-ph",
    doi = "10.1088/1475-7516/2018/10/022",
    journal = "JCAP",
    volume = "10",
    pages = "022",
    year = "2018"
}

@article{Alonso-Alvarez:2019ixv,
    author = "Alonso-{\'A}lvarez, Gonzalo and Hugle, Thomas and Jaeckel, Joerg",
    title = "{Misalignment {\textbackslash}{\&} Co.: (Pseudo-)scalar and vector dark matter with curvature couplings}",
    eprint = "1905.09836",
    archivePrefix = "arXiv",
    primaryClass = "hep-ph",
    doi = "10.1088/1475-7516/2020/02/014",
    journal = "JCAP",
    volume = "02",
    pages = "014",
    year = "2020"
}

@article{Bunch:1978yq,
    author = "Bunch, T. S. and Davies, P. C. W.",
    title = "{Quantum Field Theory in de Sitter Space: Renormalization by Point Splitting}",
    doi = "10.1098/rspa.1978.0060",
    journal = "Proc. Roy. Soc. Lond. A",
    volume = "360",
    pages = "117--134",
    year = "1978"
}

@article{Garcia:2022vwm,
    author = "Garcia, Marcos A. G. and Pierre, Mathias and Verner, Sarunas",
    title = "{Scalar dark matter production from preheating and structure formation constraints}",
    eprint = "2206.08940",
    archivePrefix = "arXiv",
    primaryClass = "hep-ph",
    reportNumber = "DESY-22-104",
    doi = "10.1103/PhysRevD.107.043530",
    journal = "Phys. Rev. D",
    volume = "107",
    number = "4",
    pages = "043530",
    year = "2023"
}

@article{Vilenkin:1982wt,
    author = "Vilenkin, Alexander and Ford, L. H.",
    title = "{Gravitational Effects upon Cosmological Phase Transitions}",
    reportNumber = "TUTP-82-7",
    doi = "10.1103/PhysRevD.26.1231",
    journal = "Phys. Rev. D",
    volume = "26",
    pages = "1231",
    year = "1982"
}

@article{Linde:1982uu,
    author = "Linde, Andrei D.",
    title = "{Scalar Field Fluctuations in Expanding Universe and the New Inflationary Universe Scenario}",
    reportNumber = "Print-82-0555 (LEBEDEV INST)",
    doi = "10.1016/0370-2693(82)90293-3",
    journal = "Phys. Lett. B",
    volume = "116",
    pages = "335--339",
    year = "1982"
}

@article{Starobinsky:1982ee,
    author = "Starobinsky, Alexei A.",
    title = "{Dynamics of Phase Transition in the New Inflationary Universe Scenario and Generation of Perturbations}",
    doi = "10.1016/0370-2693(82)90541-X",
    journal = "Phys. Lett. B",
    volume = "117",
    pages = "175--178",
    year = "1982"
}

@article{BICEP:2021xfz,
    author = "Ade, P. A. R. and others",
    collaboration = "BICEP, Keck",
    title = "{Improved Constraints on Primordial Gravitational Waves using Planck, WMAP, and BICEP/Keck Observations through the 2018 Observing Season}",
    eprint = "2110.00483",
    archivePrefix = "arXiv",
    primaryClass = "astro-ph.CO",
    doi = "10.1103/PhysRevLett.127.151301",
    journal = "Phys. Rev. Lett.",
    volume = "127",
    number = "15",
    pages = "151301",
    year = "2021"
}

@article{Turner:1987vd,
    author = "Turner, Michael S. and Widrow, Lawrence M.",
    title = "{Gravitational Production of Scalar Particles in Inflationary Universe Models}",
    reportNumber = "FERMILAB-PUB-87-194-A",
    doi = "10.1103/PhysRevD.37.3428",
    journal = "Phys. Rev. D",
    volume = "37",
    pages = "3428",
    year = "1988"
}

@article{Weinberg:2003sw,
    author = "Weinberg, Steven",
    title = "{Adiabatic modes in cosmology}",
    eprint = "astro-ph/0302326",
    archivePrefix = "arXiv",
    reportNumber = "UTTG-12-02",
    doi = "10.1103/PhysRevD.67.123504",
    journal = "Phys. Rev. D",
    volume = "67",
    pages = "123504",
    year = "2003"
}

@article{Allali:2025pja,
    author = "Allali, Itamar J. and Chakraborty, Priyesh and Fan, JiJi and Reece, Matthew",
    title = "{Axion perturbations: A general analytical treatment}",
    eprint = "2510.07371",
    archivePrefix = "arXiv",
    primaryClass = "hep-ph",
    doi = "10.1103/gwf6-1w8z",
    journal = "Phys. Rev. D",
    volume = "114",
    number = "4",
    pages = "043517",
    year = "2026"
}

@article{Linde:1996gt,
    author = "Linde, Andrei D. and Mukhanov, Viatcheslav F.",
    title = "{Nongaussian isocurvature perturbations from inflation}",
    eprint = "astro-ph/9610219",
    archivePrefix = "arXiv",
    reportNumber = "SU-ITP-96-49",
    doi = "10.1103/PhysRevD.56.R535",
    journal = "Phys. Rev. D",
    volume = "56",
    pages = "R535--R539",
    year = "1997"
}

@article{Peebles:1998nu,
    author = "Peebles, P. J. E.",
    title = "{An isocurvature cdm cosmogony. I. a worked example of evolution through inflation}",
    eprint = "astro-ph/9805194",
    archivePrefix = "arXiv",
    reportNumber = "PUPGGT98-2",
    doi = "10.1086/306619",
    journal = "Astrophys. J.",
    volume = "510",
    pages = "523",
    year = "1999"
}

@article{Langlois:2008vk,
    author = "Langlois, David and Vernizzi, Filippo and Wands, David",
    title = "{Non-linear isocurvature perturbations and non-Gaussianities}",
    eprint = "0809.4646",
    archivePrefix = "arXiv",
    primaryClass = "astro-ph",
    doi = "10.1088/1475-7516/2008/12/004",
    journal = "JCAP",
    volume = "12",
    pages = "004",
    year = "2008"
}

@article{Chung:2004nh,
    author = "Chung, Daniel J. H. and Kolb, Edward W. and Riotto, Antonio and Senatore, Leonardo",
    title = "{Isocurvature constraints on gravitationally produced superheavy dark matter}",
    eprint = "astro-ph/0411468",
    archivePrefix = "arXiv",
    reportNumber = "FERMILAB-PUB-04-328-A, MIT-CTP-3565",
    doi = "10.1103/PhysRevD.72.023511",
    journal = "Phys. Rev. D",
    volume = "72",
    pages = "023511",
    year = "2005"
}

@misc{Petretti:2026ayw,
    author = "Petretti, Catherine and Singh, Praniti and Braglia, Matteo and Chen, Xingang and Fan, JiJi and Li, Lingfeng",
    title = "{CMB Constraints on Pre-Inflationary Axion Dark Matter Isocurvature}",
    eprint = "2606.11312",
    archivePrefix = "arXiv",
    primaryClass = "astro-ph.CO",
    reportNumber = "CERN-TH-2026-136",
    month = "6",
    year = "2026"
}

@article{Kawasaki:2008sn,
    author = "Kawasaki, Masahiro and Nakayama, Kazunori and Sekiguchi, Toyokazu and Suyama, Teruaki and Takahashi, Fuminobu",
    title = "{Non-Gaussianity from isocurvature perturbations}",
    eprint = "0808.0009",
    archivePrefix = "arXiv",
    primaryClass = "astro-ph",
    reportNumber = "IPMU-08-0050, ICRR-REPORT-527",
    doi = "10.1088/1475-7516/2008/11/019",
    journal = "JCAP",
    volume = "11",
    pages = "019",
    year = "2008"
}

@article{Ellis:2021kad,
    author = "Ellis, John and Garcia, Marcos A. G. and Nanopoulos, Dimitri V. and Olive, Keith A. and Verner, Sarunas",
    title = "{BICEP/Keck constraints on attractor models of inflation and reheating}",
    eprint = "2112.04466",
    archivePrefix = "arXiv",
    primaryClass = "hep-ph",
    reportNumber = "KCL-PH-TH/2021-90, CERN-TH-2021-199, ACT-06-21, MI-HET-769,
  UMN-TH-4107/21, FTPI-MINN-21/26",
    doi = "10.1103/PhysRevD.105.043504",
    journal = "Phys. Rev. D",
    volume = "105",
    number = "4",
    pages = "043504",
    year = "2022"
}

@article{Hikage:2012be,
    author = "Hikage, Chiaki and Kawasaki, Masahiro and Sekiguchi, Toyokazu and Takahashi, Tomo",
    title = "{CMB constraint on non-Gaussianity in isocurvature perturbations}",
    eprint = "1211.1095",
    archivePrefix = "arXiv",
    primaryClass = "astro-ph.CO",
    reportNumber = "ICRR-REPORT-634-2012-23, IPMU-12-0195",
    doi = "10.1088/1475-7516/2013/07/007",
    journal = "JCAP",
    volume = "07",
    pages = "007",
    year = "2013"
}

@article{Buckley:2025zgh,
    author = "Buckley, Matthew R. and Du, Peizhi and Fernandez, Nicolas and Weikert, Mitchell J.",
    title = "{General constraints on isocurvature from the CMB and Ly-{\ensuremath{\alpha}} forest}",
    eprint = "2502.20434",
    archivePrefix = "arXiv",
    primaryClass = "astro-ph.CO",
    doi = "10.1088/1475-7516/2025/12/006",
    journal = "JCAP",
    volume = "12",
    pages = "006",
    year = "2025"
}

@article{GrillidiCortona:2015jxo,
    author = "Grilli di Cortona, Giovanni and Hardy, Edward and Pardo Vega, Javier and Villadoro, Giovanni",
    title = "{The QCD axion, precisely}",
    eprint = "1511.02867",
    archivePrefix = "arXiv",
    primaryClass = "hep-ph",
    doi = "10.1007/JHEP01(2016)034",
    journal = "JHEP",
    volume = "01",
    pages = "034",
    year = "2016"
}

@article{Borsanyi:2016ksw,
    author = "Borsanyi, Sz. and others",
    title = "{Calculation of the axion mass based on high-temperature lattice quantum chromodynamics}",
    eprint = "1606.07494",
    archivePrefix = "arXiv",
    primaryClass = "hep-lat",
    reportNumber = "DESY-16-105",
    doi = "10.1038/nature20115",
    journal = "Nature",
    volume = "539",
    number = "7627",
    pages = "69--71",
    year = "2016"
}

@article{DiLuzio:2020wdo,
    author = "Di Luzio, Luca and Giannotti, Maurizio and Nardi, Enrico and Visinelli, Luca",
    title = "{The landscape of QCD axion models}",
    eprint = "2003.01100",
    archivePrefix = "arXiv",
    primaryClass = "hep-ph",
    reportNumber = "DESY 20-036, DESY-20-036",
    doi = "10.1016/j.physrep.2020.06.002",
    journal = "Phys. Rept.",
    volume = "870",
    pages = "1--117",
    year = "2020"
}

@article{LiteBIRD:2022cnt,
    author = "Allys, E. and others",
    collaboration = "LiteBIRD",
    title = "{Probing Cosmic Inflation with the LiteBIRD Cosmic Microwave Background Polarization Survey}",
    eprint = "2202.02773",
    archivePrefix = "arXiv",
    primaryClass = "astro-ph.IM",
    doi = "10.1093/ptep/ptac150",
    journal = "PTEP",
    volume = "2023",
    number = "4",
    pages = "042F01",
    year = "2023"
}

@article{Domenech:2021ztg,
    author = "Dom{\`e}nech, Guillem",
    title = "{Scalar Induced Gravitational Waves Review}",
    eprint = "2109.01398",
    archivePrefix = "arXiv",
    primaryClass = "gr-qc",
    doi = "10.3390/universe7110398",
    journal = "Universe",
    volume = "7",
    number = "11",
    pages = "398",
    year = "2021"
}

@article{Carrasco:2015rva,
    author = "Carrasco, John Joseph M. and Kallosh, Renata and Linde, Andrei",
    title = "{Cosmological Attractors and Initial Conditions for Inflation}",
    eprint = "1506.00936",
    archivePrefix = "arXiv",
    primaryClass = "hep-th",
    doi = "10.1103/PhysRevD.92.063519",
    journal = "Phys. Rev. D",
    volume = "92",
    number = "6",
    pages = "063519",
    year = "2015"
}

@article{Kallosh:2013yoa,
    author = "Kallosh, Renata and Linde, Andrei and Roest, Diederik",
    title = "{Superconformal Inflationary $\alpha$-Attractors}",
    eprint = "1311.0472",
    archivePrefix = "arXiv",
    primaryClass = "hep-th",
    doi = "10.1007/JHEP11(2013)198",
    journal = "JHEP",
    volume = "11",
    pages = "198",
    year = "2013"
}

@article{Ellis:2019bmm,
    author = "Ellis, John and Nanopoulos, Dimitri V. and Olive, Keith A. and Verner, Sarunas",
    title = "{Unified No-Scale Attractors}",
    eprint = "1906.10176",
    archivePrefix = "arXiv",
    primaryClass = "hep-th",
    reportNumber = "KCL-PH-TH/2019-51, CERN-TH-2019-089, ACT-03-19, MI-TH-1923, UMN-TH-3826/19, MI-TH-1923,
  UMN-TH-3826/19, FTPI-MINN-19/17",
    doi = "10.1088/1475-7516/2019/09/040",
    journal = "JCAP",
    volume = "09",
    pages = "040",
    year = "2019"
}

@article{Ellis:2025zrf,
    author = "Ellis, John and Garcia, Marcos A. G. and Olive, Keith A. and Verner, Sarunas",
    title = "{Constraints on attractor models of inflation and reheating from Planck, BICEP/Keck, ACT DR6, and SPT-3G data}",
    eprint = "2510.18656",
    archivePrefix = "arXiv",
    primaryClass = "hep-ph",
    reportNumber = "UMN-TH-4512/25, FTPI-MINN-25/14, KCL-PH-TH/2025-42, CERN-TH-2025-199",
    doi = "10.1103/d35r-7bn8",
    journal = "Phys. Rev. D",
    volume = "113",
    number = "6",
    pages = "063571",
    year = "2026"
}

@article{Kallosh:2010xz,
    author = "Kallosh, Renata and Linde, Andrei and Rube, Tomas",
    title = "{General inflaton potentials in supergravity}",
    eprint = "1011.5945",
    archivePrefix = "arXiv",
    primaryClass = "hep-th",
    doi = "10.1103/PhysRevD.83.043507",
    journal = "Phys. Rev. D",
    volume = "83",
    pages = "043507",
    year = "2011"
}
\end{document}